\documentclass{article}

\usepackage{microtype}
\usepackage{graphicx}
\usepackage{booktabs} 

\usepackage{hyperref}

\usepackage[accepted]{icml2024}

\usepackage{amsmath}
\usepackage{amssymb}
\usepackage{mathtools}
\usepackage{amsthm}

\usepackage{stmaryrd}
\usepackage{subcaption}
\usepackage{multirow}
\usepackage{amsthm}
\usepackage{xcolor}

\usepackage{xspace}
\newcommand{\hqcom}[1]{\textcolor{black}{#1}}
\newcommand{\camera}[1]{\textcolor{black}{#1}}
\newcommand{\name}{\texttt{Ditto}\xspace}
\newcommand{\spu}{\texttt{SecretFlow-SPU}\xspace}

\newcommand{\share}[1]{\ensuremath{\llbracket #1\rrbracket}\xspace}
\newcommand{\twoshare}[1]{\ensuremath{\langle #1\rangle}\xspace}

\usepackage[capitalize,noabbrev]{cleveref}

\theoremstyle{plain}
\newtheorem{theorem}{Theorem}[section]

\theoremstyle{definition}

\theoremstyle{remark}

\usepackage[textsize=tiny]{todonotes}

\icmltitlerunning{Ditto: Quantization-aware Secure Inference of Transformers upon MPC}

\begin{document}

\twocolumn[
\icmltitle{Ditto: Quantization-aware Secure Inference of Transformers upon MPC}



\icmlsetsymbol{equal}{*}

\begin{icmlauthorlist}
\icmlauthor{Haoqi Wu}{comp}
\icmlauthor{Wenjing Fang}{comp}
\icmlauthor{Yancheng Zheng}{comp}
\icmlauthor{Junming Ma}{comp}
\icmlauthor{Jin Tan}{comp}
\icmlauthor{Yinggui Wang}{comp}
\icmlauthor{Lei Wang}{comp}
\end{icmlauthorlist}

\icmlaffiliation{comp}{Ant Group, Hangzhou, China}

\icmlcorrespondingauthor{Haoqi Wu}{haoqi.whq@antgroup.com}

\icmlkeywords{Machine Learning, ICML}

\vskip 0.3in
]



\printAffiliationsAndNotice{}  

\begin{abstract}
Due to the rising privacy concerns on sensitive client data and trained models like Transformers, secure multi-party computation (MPC) techniques are employed to enable secure inference despite attendant overhead. Existing works attempt to reduce the overhead using more MPC-friendly non-linear function approximations. However, the integration of quantization widely used in plaintext inference into the MPC domain remains unclear. 
To bridge this gap, we propose the framework named \name to enable more efficient quantization-aware secure Transformer inference.
Concretely, we first incorporate an MPC-friendly quantization into Transformer inference and employ a quantization-aware distillation procedure to maintain the model utility. Then, we propose novel MPC primitives to support the type conversions that are essential in quantization and implement the quantization-aware MPC execution of secure quantized inference.
This approach significantly decreases both computation and communication overhead, leading to improvements in overall efficiency.
We conduct extensive experiments on Bert and GPT2 models to evaluate the performance of \name. The results demonstrate that \name is about $3.14\sim 4.40\times$ faster than MPCFormer (ICLR 2023) and $1.44\sim 2.35\times$ faster than the state-of-the-art work PUMA with negligible utility degradation.
\end{abstract}

\section{Introduction}\label{sec:intro}
The recent advance of pre-trained Transformer~\cite{attention-all-need-17} models in domains like visual recognition~\cite{vit-21, slef-supervised-vit-ICCV21} and natural language processing~\cite{bert-19, gpt2-19} have led to their widespread adoption for machine learning (ML) inference services. However, despite their increasing popularity, a major concern revolves around data security.
In ML services like ChatGPT~\cite{chatgpt}, the model owner offers an API that receives user prompts as input and generates answers in return. However, this process involves sending user prompts to the server in plaintext, potentially exposing sensitive information.
An alternative approach is to employ secure multi-party computation (MPC)~\cite{shamir-79, yao-mpc} to offer provable security.

However, the huge computation and communication overhead introduced by MPC techniques hinders the application of MPC-based secure Transformer inference. For one thing, non-linear functions like GeLU are frequently invoked, which are expensive in MPC.
For another, the overhead is amplified in Transformers due to their large model size. In general, Transformers have millions to billions of parameters. 
As to the former problem, \cite{faster-CryptoNets, li2023mpcformer, privformer, liu2023llms} replace these non-linear functions with MPC-friendly approximations.
Regarding the latter, there have been practices in plaintext inference~\cite{llm.int8-22, i-bert-icml21} that quantize the model to lower bits and employ low-bit integer arithmetic, thus accelerating the inference.
However, plaintext quantization \textbf{cannot be trivially incorporated} into secure inference upon MPC.

\begin{figure}[ht]
    \centering
    \includegraphics[scale=0.35]{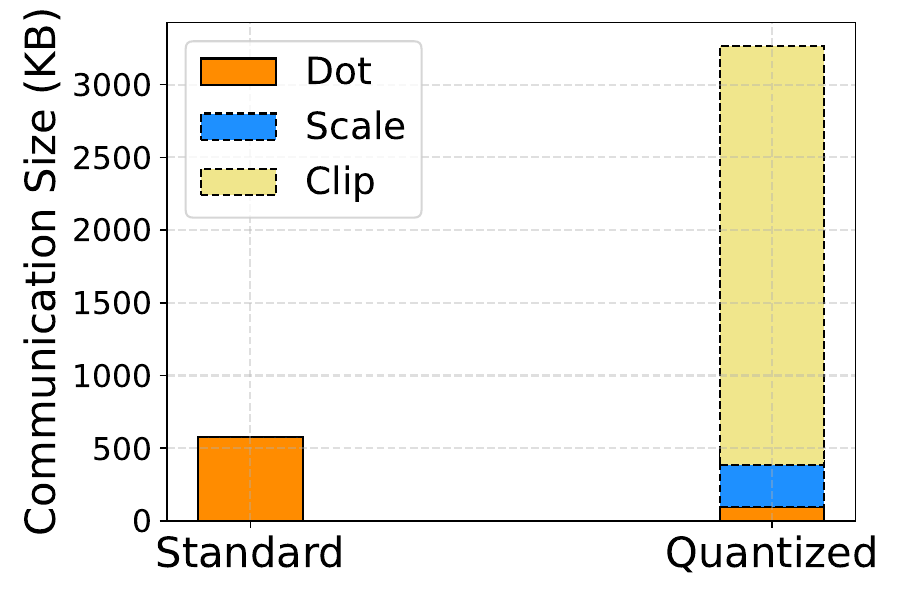}
    \caption{Communication size breakdown on standard vs. quantized secure matrix multiplication. 
    $\mathbf{Y} = \mathbf{X}^{8\times 768} \cdot \mathbf{W}^{768\times 3072}$. Scale denotes multiplying the dot product by the scaling factor $s$. Clip typically restricts the values within the range $[-128, 127]$ for int8 quantization. 
    }
    \label{fig:matmul-example}
\end{figure}

\hqcom{\noindent\textbf{Toy Example.} For a linear layer in Transformers $\mathbf{Y} = \mathbf{X} \cdot \mathbf{W}$, $\mathbf{Y} \in \mathbb{R}^{B \times M}$, $\mathbf{X} \in \mathbb{R}^{B\times N}$ and $\mathbf{W} \in \mathbb{R}^{N \times M}$, where $B$ is the batch size, $N$ and $M$ are hidden dimensions. 
A standard floating-point matrix multiplication over MPC typically operates over uniform 64-bit fixed-point numbers $\mathbf{Y}_{\text{fxp}64}= \mathbf{X}_{\text{fxp}64} \cdot \mathbf{W}_{\text{fxp}64}$ , while quantized matrix multiplication goes as $\mathbf{Y}_{i8}=\lfloor (\mathbf{X}_{i8} \cdot \mathbf{W}_{i8})_{i32} \cdot s \rceil$, where the multiply-and-accumulate of $\mathbf{X}_{i8} \cdot \mathbf{W}_{i8}$ is computed over 32-bit integers, $s$ is a floating-point scaling factor and $\lfloor \cdot \rceil$ denotes the clip operation. 
As demonstrated in Figure~\ref{fig:matmul-example}, while quantization indeed diminishes the bitwidth, thereby reducing the communication of dot product, the additional procedures, i.e., scaling and clipping impose substantial overhead, surpassing the benefits derived from quantization.}

It is worth noting that there are several cross-domain gaps between the worlds of ML and MPC since

\begin{itemize}
    \setlength{\itemsep}{0pt}
    \item \textbf{ML experts} mainly focus on designing delicate quantization methods to improve efficiency, which, however, may not be MPC-friendly. The type conversions between data types like \texttt{INT8} and \texttt{FP16} in plaintext quantization that involves scaling and clipping operations are not trivial in MPC. 
    Besides, applying quantization directly may cause substantial utility drop.
    \item \textbf{MPC experts} mainly focus on constructing efficient underlying primitives and may not be aware of employing mixed-precision quantization to enhance end-to-end inference efficiency, thus lacking the capability to support quantization-aware secure inference.
\end{itemize}

Therefore, the problem naturally arises: 

\begin{center}
    \textit{Can we perform quantization-aware secure inference with negligible utility degradation?}
\end{center}

As an answer to the above question, we develop \name, an \textit{easy-to-deploy} framework for secure and efficient Transformer inference based on a co-design of novel MPC primitives and ML quantization. Our contributions are as follows:

\begin{itemize}
    \setlength{\itemsep}{0pt}
    \item \textbf{MPC-friendly Quantization-Aware Distillation}. We propose to incorporate \textit{static dyadic quantization} (i.e., from floating-point to fixed-point representation) to avoid CPU-cheap yet MPC-expensive dynamic quantization operations like \texttt{clip} and \texttt{max}. With lower quantization precision, a smaller bitwidth is required, thus reducing the overall overhead in secure inference. Besides, we utilize knowledge distillation to perform \textit{quantization-aware distillation} on pre-trained models to retain the model utility. 
    \item \textbf{Secure Quantized Inference Framework.} To the best of our knowledge, \name is the \textit{first} framework that supports the MPC execution of quantization-aware secure inference. Concretely, the layer-wise quantized computations are automatically mapped to secure computations over different data types (essentially different rings). To do so, we propose novel MPC primitives to support efficient interchangeable type conversions. We will open-source the code once accepted.
    \item \textbf{Empirical Evaluations.} We evaluate \name in the secure inference of several Transformer models, i.e., Bert and GPT2. The performance is evaluated from two metrics: the model utility and efficiency. The evaluation results indicate that efficiency improvement can be achieved without a significant utility drop. Compared to prior works, \name is about $3.14\sim 4.40\times$ faster than MPCFormer~\cite{li2023mpcformer} and $1.44\sim 2.35\times$ faster than PUMA~\cite{puma-2023}.
\end{itemize}


\section{Related Work}\label{sec:related-work}
\paragraph{Secure Transformer Inference.}

MPC originates from the Billionaire problem~\cite{yao-mpc, shamir-79} and aims to enable multiple untrusted parties to jointly compute a function while keeping the inputs private. There have been many prior works working on privacy-preserving machine learning (mainly focusing on convolutional neural networks), including \textbf{two-party computation (2PC) setting}~\cite{secureml-17, aby2.0-21, cheetah-22}, \textbf{3PC setting}~\cite{aby3, cryptgpu-sp21, falcon-20} and even \textbf{4PC setting}~\cite{flash-20, fantastic-four-21}. 
Recent works~\cite{li2023mpcformer, hao2022iron, privformer, liang2023merge, liu2023llms, puma-2023, SAL-ViT-23} further study the secure inference of more complex Transformer models. 
These approaches mainly use MPC-friendly approximations for non-linear functions. 
\camera{For example, MPCFormer~\cite{li2023mpcformer}, MPCViT~\cite{mpcvit-23} and SAL-ViT~\cite{SAL-ViT-23} propose more efficient \textit{softmax} approximations, which are complementary to our work. We take the first step towards leveraging MPC-friendly quantization to enhance the efficiency.}
Among these works, \textbf{3PC in semi-honest and honest-majority setting}~\cite{li2023mpcformer, puma-2023} achieves the overall best efficiency. In this work, we also adopt this setting.

\hqcom{\noindent\textbf{Secure Quantized Inference.}
In MPC-based secure inference, floating-point arithmetic is typically quantized to fixed-points to enable integer computations over rings~\cite{li2023mpcformer, puma-2023, aby3, secureq8}. 
XONN~\cite{XONN} and Quotient~\cite{quotient} utilize binary or ternary quantized weights to accelerate inference. Nonetheless, such quantization may result in utility drop, especially in large Transformers.
SecureQ8~\cite{secureq8} pioneers the use of int8 quantization in secure inference. However, the quantization of plaintext weights/activations to 8-bit integers does not extend to ciphertexts, which remain encoded over a uniformly large ring, typically 64-bit $\mathbb{Z}_{2^{64}}$. This discrepancy stems from the absence of type conversion methods and the necessity for higher bitwidths in certain non-linear computations to maintain sufficient precision. This inconsistency impedes the potential for efficiency gains that SecureQ8 might otherwise offer. In this paper, we aim to resolve the above drawbacks for secure Transformer inference.}

\section{Background}\label{sec:background}
In this section, we briefly introduce the Transformer models and the underlying MPC protocols, specifically 2-out-of-3 replicated secret sharing.
Finally, we introduce the quantization methods, along with the fixed-point quantization.



\subsection{Transformer and its Variants}\label{sec:transformers}
Transformer models generally consist of three parts: 1) the \texttt{Embedding} module that maps a discrete token to its continuous hidden vector representation; 2) a stack of Transformer \texttt{Block}; 3) the last \texttt{Prediction} module that maps the hidden vector to task-specific representation. For Transformer \texttt{Block}, it typically has \texttt{Attention} and \texttt{Feed-Forward Network (FFN)} modules.

\texttt{Attention} module can be formulated as $\mathsf{Softmax}(\mathsf{Q}\cdot \mathsf{K}^\top+\mathsf{M})\cdot \mathsf{V}$, where $\mathsf{Q}, \mathsf{K}, \mathsf{V}$ denote the vectors obtained by the matrix multiplication of input activation and three weight matrices, and $\mathsf{M}$ denotes the attention mask. The two widely-used variants, Bert and GPT, use different masks. 

\texttt{FFN} module can be formulated as $\mathsf{Linear}(\mathsf{GeLU}(\mathsf{Linear}(x, w_0, b_0)), w_1, b_1)$, where $w_i, b_i$ denote the parameters for $i$-th linear layer. It consists of two linear layers and an activation function $\mathsf{GeLU}$.

\subsection{Secure Multi-Party Computation}\label{sec:mpc-intro}
2-out-of-3 replicated secret sharing (RSS)~\cite{rss-16, aby3}, a widely-used MPC technique, runs by splitting a secret value $x$ into several random values (denoted as shares) as $\share{x} = \{x_0$, $x_1$, $x_2\}$, s.t., $x = x_0 + x_1 + x_2 \mod 2^\ell$, where $\ell$ denotes the ring size. All the computations are performed over the ring $\mathbb{Z}_2^\ell$. In RSS, the three shares are distributed to three computing parties $\mathcal{P} = \{P_0, P_1, P_2\}$, where $P_i$ holds two shares $\{x_i, x_{i+1}\}$ ($x_3$ corresponds to $x_0$).

In this paper, we use $[\![\cdot]\!]_\ell$ to denote RSS over $\mathbb{Z}_{2^\ell}$. For $\ell \ge 1$ that supports arithmetic operations like $+, -, \cdot$, we denote such type as \textit{arithmetic sharing}. In the case of $\ell = 1$ that only supports boolean operations like bit-wise $\oplus$ and $\land$, we refer to this type as \textit{boolean sharing}. 

In order to incorporate floating-point arithmetic, which is extensively used in ML, into MPC that operates over a ring, we employ fixed-point quantization to encode floating-point numbers as integers. This approach can be considered as a branch of quantization techniques (refer to Section \ref{sec:quantization}).


\textbf{Linear Algebra.}
Let $a, b, c$ be public constants and $\share{x}, \share{y}$ be arithmetic-shared values. $a\share{x}+b\share{y}+c$ only involves addition and multiplication by a public constant. Hence, $\share{ax + by + c}$ can be computed as $(ax_0+by_0+c, ax_1+by_1, ax_2+by_2)$ locally.
While for the multiplication of two shared values, $\share{x\cdot y}$ can be decomposed into $(x_0+x_1+x_2)\cdot (y_0+y_1+y_2) = \sum_{i=0}^2z_i=(x_iy_i+x_{i+1}y_i+x_iy_{i+1})$, with $P_i$ computes $z_i$. To obtain $\share{x\cdot y}$, the parties should perform \textit{re-share} $z_i$~\cite{aby3}, which requires communication with each other. 

\textbf{Non-linear Functions.}
In addition to linear algebra operations, non-linear functions such as Softmax and Layer Normalization are commonly employed. To implement these functions, we leverage several underlying MPC computation primitives proposed in prior works~\cite{aby3}. We omit the descriptions for primitives like comparison, which are used as black boxes, and only present the functionalities of primitives explicitly mentioned in this paper.
We refer to \cite{puma-2023, rsqrt} to construct ${\mathsf{Exp}}(\share{x}) = \share{e^x}$ and ${\mathsf{rSqrt}}(\share{x}) = \share{1/\sqrt{x}}$.

\subsection{Model Quantization}\label{sec:quantization}
Quantization~\cite{quantization-survey-21} refers to converting floating-point numbers to low-bit integer representation like 8-bit integer. This can be formulated as $\hat{x} = \mathsf{Clip}(\mathsf{Int}(x/s), min, max)$, where $min,max$ denote the clipping bound and $s$ denotes the scaling factor. Generally, quantization methods can be divided into two categories: post-training quantization (PTQ)~\cite{ZeroQuant-nips22, llm.int8-22, OPTQ-23} and quantization-aware training (QAT)~\cite{i-bert-icml21, hawqv3-icml21}. The former PTQ methods allow one-shot quantization while requiring more complex quantization schemes, e.g., more fine-grained quantization~\cite{ZeroQuant-nips22} or dynamic quantization~\cite{llm.int8-22}.
The latter QAT methods allow for more diverse quantization by quantizing the weights and activations during the training of the model. Hence, cheaper quantization methods like static quantization are feasible despite the cost of \textit{re-training} the model.

Most of the quantization methods utilize floating-point scales to achieve adequate precision. Among these works, \textit{dyadic} quantization~\cite{hawqv3-icml21, i-bert-icml21} is a typical class for integer-only quantization, where the scale $s$ is a dyadic number $c/2^{f}$, $c$ is an integer and $f$ is the precision bit.
In this paper, we employ a modified version called fixed-point quantization ($s=1/2^f$, with $c=1$) to accommodate floating-point computations into fixed-point arithmetic, which is crucial in MPC.
Mathematically, this can be expressed as $\hat{x} = \texttt{FXP}_\ell^f(x) = \lfloor x*2^f \rceil \mod 2^\ell$.

\section{Design}\label{sec:design}
In this section, we begin by introducing the high-level workflow of \name. Then we elaborate on two ingredients in \name: 1) the MPC-friendly quantization and distillation of Transformers; 2) the quantization-aware secure inference of the quantized and distilled model upon MPC.

\subsection{High-Level Workflow}\label{sec:workflow}
\noindent\textbf{Setting.} In this paper, we consider the inference scenario, where the model owner provides a trained model $\mathcal{M}$, and the client provides input data $x$ for the inference task. 
The inference computation can be formulated as $y = \mathcal{M}_\theta(x)$, where $\theta$ denotes the parameters for the model $\mathcal{M}$.
The security concern is that both the parameters $\theta$ and input data $x$ are unknown to each other, along with potential attackers. Only the inference result $y$ can be revealed to the client.

Similar to prior works~\cite{li2023mpcformer, puma-2023}, we consider the secure outsourced 3PC setting. That is, we offload the inference computation to an MPC system consisting of three computing parties $\mathcal{P} = \{P_0, P_1, P_2\}$. The client encrypts $\share{x}$ using RSS and sends the shares to corresponding computing parties. Similarly, the model owner encrypts the model parameters $\theta$ and sends $\share{\theta}$ to $\mathcal{P}$. The computing parties $\mathcal{P}$ then carry out the secure inference and obtain the inference result $\share{y}$. $\mathcal{P}$ then sends all the shares of $y$ to the client, who can reveal the plaintext of $y$.

\noindent\textbf{Security Model.} In the MPC system, we consider \textit{semi-honest} and \textit{honest-majority} adversary~\cite{puma-2023, cryptgpu-sp21}, where the adversary corrupts no more than half of the computing parties (one exactly in 3PC) and strictly follow the protocol to perform computations but might try to crack sensitive information by analyzing the messages they receive. We note that output privacy, where the inference outputs can be utilized to infer information like membership~\cite{membership-inference-17}, is beyond the scope.


\begin{figure}[h]
    \centering
    \includegraphics[scale=0.39]{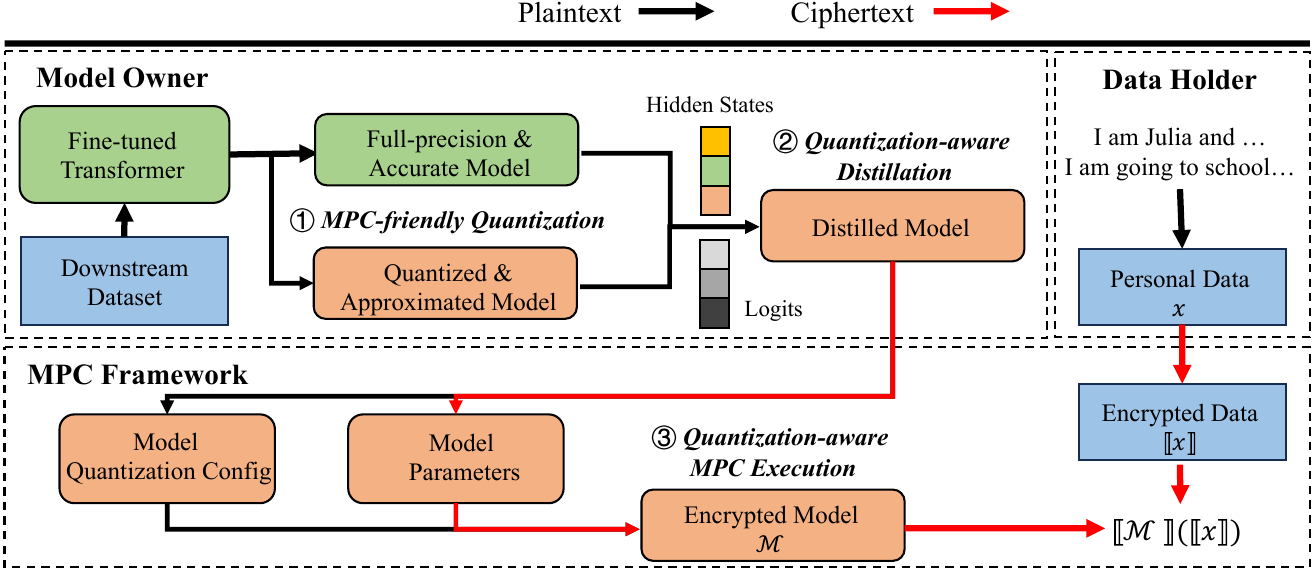}
    \caption{High-level workflow of \name.}
    \label{fig:workflow}
\end{figure}

With the setting and security model in mind, we hereby present the high-level workflow of \name in Figure~\ref{fig:workflow}.
In general, this is a two-step inference scheme via a co-design of ML quantization and efficient MPC computation. 
The first step (the \textbf{upper left} part) is quantizing and distilling the model to a more MPC-friendly version (Section~\ref{sec:mpc-friendly-qd}). This step is performed by the model owner locally using plaintext computation. 
The second step (the \textbf{bottom} part) involves quantization-aware secure inference of the MPC-friendly model obtained from the first step. We design novel MPC primitives to support essential type conversion in quantization (Section~\ref{sec:mpc-infer}). 

\subsection{MPC-friendly Model Quantization and Distillation}\label{sec:mpc-friendly-qd}

\subsubsection{MPC-friendly Fixed-point Model Quantization}\label{sec:fixed-point-arithmetic}

Our motivation to employ quantization in secure inference originates from the following two considerations.
1) The necessity of model quantization is amplified in secure inference 
of Transformers since it requires communicating messages between computing parties and the communication size heavily depends on the message bitwidth. Therefore, applying quantization to decrease bitwidth can theoretically shrink the overall volume of transmitted data, leading to improved inference efficiency.
2) The feasibility of using low-bit quantization is also evidenced in previous works. As observed in~\cite{over-parameterization-nips22}, the neural network models are typically over-parameterized, thus leaving room for reducing the precision and computing with lower bitwidth while maintaining the model utility. 
The recent success of quantization in large language models~\cite{llm.int8-22, GPTQ-22} also proves the feasibility of quantization in more complex models.

Whereas there exist several gaps between plaintext quantized inference and secure quantized inference.

\textbf{\textit{Gap 1: Dynamic quantization is expensive in MPC.}}
The state-of-the-art plaintext quantization works~\cite{llm.int8-22, OPTQ-23} allow the entire inference to be carried out using low-bit integers like \texttt{INT8} or even \texttt{INT4}.
Despite achieving considerable speed up, tailored quantization operations are required, like dynamically computing the min/max/outlier to obtain the scaling factor $s$ and calculating $\lfloor x \cdot s \rceil$ with clipping. Such operations are quite expensive in MPC.
In this case, directly transplanting existing quantization strategies into secure inference is not viable, where the overhead to perform quantization alone outweighs the overhead reduction brought by quantization. 

\textbf{Solution.} To mitigate this issue, we adopt \textbf{static dyadic quantization} to avoid dynamically computing scale in inference.
By dyadic, we quantize all the weights and activations into fixed-point numbers, where the scaling factor is in the form of $1/2^{f}$. In this way, the scale conversion can be implemented using left-shift or right-shift (aka. truncation~\cite{aby3, mixed-ab-circuit-20}), which is much cheaper in MPC.
We also adopt layer-wise quantization. That is, we use different quantization scales for different layers. By enabling a smaller quantization scale for linear layers that are not sensitive to precision, we can improve efficiency. While for those non-linear layers like layer normalization, we use a larger scale to avoid a significant accuracy drop.~\cite{mp-training-19}~\footnote{Similar 
 precision practices are also used in PyTorch: \url{https://pytorch.org/docs/stable/amp.html\#cuda-ops-that-can-autocast-to-float16}}.

\textbf{\textit{Gap 2: Type conversions are difficult in MPC.}}
In plaintext quantization, linear operations utilize \texttt{INT8} for computation and \texttt{INT32} for accumulation to prevent overflow, while non-linear functions compute in \texttt{FP32} to maintain sufficient precision. This necessitates type conversions throughout the model inference to avert overflow and substantial loss of precision.
However, we note that these type conversions between \texttt{INT8} and \texttt{INT32} are straightforward in plaintext but pose a novel challenge in MPC, which operates over rings. Such type conversion involves converting shares among different rings, which cannot be done locally. 

\textbf{Solution.} To bridge this gap, we propose efficient \textbf{type conversion MPC primitives} (Section~\ref{sec:mpc-protocol}). 
Besides, to avoid frequent type conversions, we use uniform low-bitwidth fixed-point encoding ($\texttt{FXP}_{32}^8$) in intermediate Transformer blocks for linear operations.
For the non-linear functions and the last prediction layer, we instead use a higher-bitwidth fixed-point encoding ($\texttt{FXP}_{64}^{18}$) for the sake of adequate precision.
We note that 32/64-bit integers can accommodate the activation range, of which the distribution is illustrated in Appendix~\ref{append:distribution}, thus avoiding overflow.

To summarize, we quantize all the weights and activations into fixed-point numbers using layer-wise static quantization with dyadic scales. 
In \name, all the variables (i.e., activations and weights) are quantized with different precision. 
We provide an illustration of the difference between fixed-point inference and traditional floating-point inference in Appendix~\ref{append:fixed-point-inference}.
Note that for the non-linear layers, we exploit  \textbf{fixed-point conversion} to raise the precision. 


\subsubsection{Quantization-aware Activation Functions}\label{sec:qa-functions}
We hereby introduce the \textbf{quantization-aware} non-linear activation functions using fixed-point numbers of different bits.
The formulated algorithms are deferred to Appendix~\ref{append:protocols}.

\textbf{Softmax.} We use Softmax to map the inputs into the range $(0, 1)$ as $\mathsf{Softmax}(x) = \frac{e^{x_i}}{\sum_i e^{x_i}}$. For numerical stability, we first `normalize' the input by computing $x = x - \mathsf{max}_i(x)$. Since $\mathsf{max}$ does not require a high precision, we compute this part using $\texttt{FXP}_{32}^{8}$. For the following exponential and division, we use existing protocols ${\mathsf{Exp}}$ (tailored approximation for Softmax, detailed in Appendix~\ref{append:approximation}) and ${\mathsf{Recip}}$~\cite{catrina-10}.
These two functions are computed over $\texttt{FXP}_{64}^{18}$ to maintain adequate precision. 




\textbf{GeLU.} The original GeLU function computes $\mathsf{GeLU}(x) = \frac{x}{2}\cdot (1+\tanh(\sqrt{2/\pi}\cdot(x+0.044715\cdot x^3)))$. To precisely compute GeLU, we first try to use high-order chebyshev polynomial to approximate $\tanh$. However, it requires several multiplication, thus leading to significant overhead. Inspired by~\cite{li2023mpcformer}, we use a quantized polynomial to directly approximate $\mathsf{GeLU}(x) = 0.125x^2+0.25x+0.5$ over $\texttt{FXP}_{32}^{8}$. 
The Quad approximation is worth mentioning as it evaluates a two-order polynomial, allowing it to be computed with lower precision.





\textbf{Layer Normalization.} 
Given a vector of $x$ as input, $\mathsf{LayerNorm} = \frac{x - \mu}{\sqrt{\sigma + \epsilon}}\cdot \textbf{g} + \textbf{b}$, where $\mu$ and $\sigma$ denote mean and variance, $\textbf{g}, \textbf{b}$ denote scale and bias, and $\epsilon$ is a small constant. To avoid significant precision loss, we upcast the inputs and perform the layer normalization with a relatively higher precision $\texttt{FXP}_{64}^{18}$. The final outputs are downcasted back to $\texttt{FXP}_{32}^{8}$ for subsequent computations. 

\subsubsection{Quantization-aware Distillation} 
Despite the efficiency gain from the above MPC-friendly quantization and approximation, these two steps can cause the precision drop. We illustrate the loss between the outputs from the original model and the quantized and approximated model in Appendix~\ref{append:loss-distill}.
Consequently, the converted model $\mathcal{M}$ is of low utility.
In order to compensate for the error introduced by these two methods, we adopt \textit{knowledge distillation} (KD)~\cite{tinybert-emnlp20, li2023mpcformer}.


Without special declaration, we denote the original model as $\mathcal{T}$ and the converted model as $\mathcal{M}$. 
We quantize the weights of $\mathcal{T}$ to low-precision version in layer-wise granularity, which serve as the initial weights of $\mathcal{M}$.
All the computations in $\mathcal{M}$ use integer-only arithmetic.
We leverage layer-wise distillation, considering that we use layer-wise quantization. Concretely, we capture the hidden states of all the Transformer layers from both $\mathcal{T}$ and $\mathcal{M}$ and use the Mean Squared Error loss (MSE) between these two outputs to measure the distillation loss. 
 

\subsection{Secure Model Inference upon MPC}\label{sec:mpc-infer}
To perform the aforementioned type conversions in layer-wise quantization, we propose efficient MPC primitives. 

\subsubsection{Type Conversion MPC primitives}\label{sec:mpc-protocol}
The type conversion can be divided into \textit{upcast} and \textit{downcast}. Upcast refers to converting values from a smaller fixed-point representation to a larger fixed-point representation, while downcast is the opposite.
In MPC, type conversions additionally involves share conversions among different rings. 
We consider convert the input $\share{x}_\ell$ from $\texttt{FXP}_\ell^f$ to $\texttt{FXP}_{\ell'}^{f'}$. 
Due to page limitation, we defer the correctness analysis and security proof to Appendix~\ref{append:protocols}$\sim$\ref{append:security-proof}.


\textbf{Downcast ($\ell > \ell', f > f'$, Algorithm~\ref{protocol:cast-down}).} It suffices to a right-shift followed by a modulo operation as $\share{x'}_{\ell'} = \mathsf{DownCast}(\share{x}_{\ell}) = x_i \gg (f - f') \mod 2^{\ell'}$ for $i \in \{0, 1, 2\}$. 
The local right-shift by $(f-f')$ bits first lowers the precision to $2^{f'}$. The subsequent local modulo operation, i.e., dropping the most significant $(\ell-\ell')$ bits, converts the shares to a smaller ring, s.t., $x/2^{f} = x'/2^{f'}$. 







\textbf{Upcast ($\ell < \ell', f < f'$, Algorithm~\ref{protocol:cast-up}).} It suffices to convert $\share{x}$ from $\mathbb{Z}_{2^\ell}$ to $\mathbb{Z}_{2^{\ell'}}$, followed by a left-shift operation. The left-shift can be implemented directly by left-shifting the shares locally. While for the ring conversion, it is not trivial. As shown in Equation~\ref{eq:upcast}, there may be potential wrap $w$ of the sum of $x_i$ modulo $2^\ell$, i.e., $w = \lfloor (x_0+x_1+x_2)/ 2^\ell \rfloor$. $w$ cannot be implicitly erased since $\ell < \ell'$ and $w \cdot 2^\ell \mod 2^{\ell'}$ might not equal zero. However, directly computing $w$ that involves non-linear comparisons is expensive in MPC. 

\begin{equation}\label{eq:upcast}
\small
\begin{split}
    x &\mod 2^\ell = (x_0 + x_1 + x_2 - w \cdot 2^\ell) \mod 2^\ell \\
    &= x_0 \mod 2^{\ell'} + x_1 \mod 2^{\ell'} + x_2 \mod 2^{\ell'} \\
    &- (w \cdot 2^\ell) \mod 2^{\ell'} \\
\end{split}
\end{equation}

\begin{algorithm}[h]
\caption{Secure $\mathsf{UpCast}$ Protocol}\label{protocol:cast-up}
\small
\begin{algorithmic}[1]
\REQUIRE
RSS-shared $\share{x}_\ell$ over $\texttt{FXP}_{\ell}^{f}$;
\ENSURE
RSS-shared $\share{x'}_{\ell'}$ over $\texttt{FXP}_{\ell'}^{f'}$, where $x'=x$.

\STATE $P_2$ samples bits $\{r_i\}$ for $i \in [0, \ell-1]$ and computes $r=\sum_{i=0}^{\ell-1}r_i*2^i$.
\STATE $P_2$ generates 2-out-of-2 sharing of $r$ and $r_{\ell-1}$ as $\twoshare{r}_{\ell'} = \{r_0, r_1\}$, $\twoshare{r_{\ell-1}}_{\ell'} = \{r_{\ell-1, 0}, r_{\ell-1, 1}\}$.

\STATE $P_2$ sends the corresponding shares to $P_0$ and $P_1$. $P_i$ holds $r_i$ and $r_{\ell-1, i}$. 

\STATE $P_i$ for $i \in \{0, 1, 2\}$ generate random numbers $z_0, z_2 \in \mathbb{Z}_{2^{\ell'}}$ using \textsf{PRF}: $P_2$ and $P_0$ samples $z_0$; $P_2$ and $P_1$ samples $z_2$.

\STATE $P_0$ and $P_1$ obtain  $\twoshare{r}_\ell$ by invoking ${\mathsf{DownCast}}(\twoshare{r}_{\ell'} \ll (f'-f))$. $~~~~~~\vartriangleright\text{using $\twoshare{r}_{\ell'}$ from Step-3}$
\STATE $P_0$ and $P_1$ convert $\share{x}_\ell$ to $\twoshare{\hat{x}}_\ell = \{\hat{x}_0, \hat{x}_1\}$ by constructing $\hat{x}_0 = x_1 + x_2$, $\hat{x}_1 = x_0$, where $P_i$ holds $\hat{x}_i$ and $x = \hat{x}$.
\STATE $P_0$ and $P_1$ executes the following steps:
\begin{equation*}
\begin{split}
&\twoshare{y}_\ell = \twoshare{\hat{x}}_\ell + \twoshare{r}_\ell\text{ and open $y = {\mathsf{Reveal}}(\twoshare{y}_\ell)$},~~~\vartriangleright y = \hat{x} + r\\
&\twoshare{\hat{w}}_{\ell'} = \twoshare{r_{\ell-1}}_{\ell'} \cdot \neg y_{\ell-1},~~~\vartriangleright \hat{w} = r_{\ell-1} \cdot \neg y_{\ell-1} \\
&\twoshare{x'}_{\ell'} = y - \twoshare{r}_{\ell'} + \twoshare{\hat{w}}_{\ell'} \cdot 2^\ell,~~~~~~\vartriangleright x = y - r + \hat{w} \cdot 2^\ell
\end{split}
\end{equation*}
and outputs $\twoshare{x'}_{\ell'} = \{x'_0, x'_1\}$, s.t., $x'_0 + x'_1 \mod 2^{\ell'} = x$.

\STATE $P_i$ for $i \in \{0, 1\}$ proceed using randomness from Step-4: 1) $P_0$ computes $x'_0 = x'_0 - z_0$; 2) $P_1$ computes $x'_1 = x'_1 - z_2$.

\STATE $P_0$ and $P_1$ exchanges $x'_0$ and $x'_1$. 
\STATE $P_0$ outputs $(z_0, x'_0 + x'_1)$, $P_1$ outputs $(x'_0 + x'_1, z_2)$ and $P_2$ outputs $(z_2, z_0)$.

\RETURN$\share{x'}_{\ell'} = \{z_0, x'_0 + x'_1, z_2\}$.

\end{algorithmic}
\end{algorithm}

\noindent\textit{Optimization Trick.} We here take the intuition of mask-and-open that goes as $x = ((x+r)\mod 2^\ell + \hat{w}\cdot 2^\ell - r) \mod 2^{\ell'}$, where $\hat{w} = (x+r) \overset{?}{>} 2^\ell$. The problem now reduces to compute another potential wrap $\hat{w}$, which is easier than computing $w$. 
\hqcom{To further facilitate the computation, we exploit a positive heuristic trick. That is, supposing the input $x \in [-2^{\ell-2}, 2^{\ell-2}-1]$ of which the MSB is unknown, we add a large bias $2^{\ell-2}$ to $x$ to ensure $x' = x + r \in [0, 2^{\ell-1}-1]$ is positive.
\camera{Take $\ell = 8$ as an example. An 8-bit signed integer can represent the range $[-128, 127]$. 
We here assume the range of $x$ as $x \in [-64, 63]$, which can be represented with a 7-bit integer. With this assumption, we have $x+64 \in [0, 127]$. This ensures that the MSB of $x+64$, when encoded as an 8-bit integer, will always be 0.}
As a result, we can compute $\hat{w} = r_{\ell-1} \wedge \neg y_{\ell-1}$ with one AND operation, where $y = x' +r \mod 2^\ell$.}
After the upcast conversion, the bias can be directly subtracted to eliminate its influence.
Noted that the sharing of $r$ over both $\mathbb{Z}_{2^{\ell}}$ and $\mathbb{Z}_{2^{\ell'}}$ (Step-5) can be implemented using $\mathsf{DownCast}$. 

\noindent{\textit{Communication Complexity Analysis.}} \hqcom{The upcast protocol in total requires sending $3\ell + \ell'$ bits in 3 communication rounds.  In Step-3, $P_2$ sends randomness of $2\ell$ bits. In Step-7, $P_0$ and $P_1$ communicates $\ell$ bits to reveal $y$. Finally, $P_0$ and $P_1$ exchanges $\ell'$ bits to obtain the output.}

\subsubsection{Quantization-aware MPC execution}
With the underlying MPC primitives ready, we proceed to implement an end-to-end secure quantization-aware inference framework. 
We enhance the \spu~\cite{spu-atc23} framework, which supports the compilation of front-end models into privacy-preserving versions but lacks support for quantization.
\camera{Concretely, we design and implement the quantization-aware compiler to allow dynamic ring support and automatic type conversions. The compiler can directly load the models from Huggingface and automatically rewrite the directed acyclic graph (DAG) to allow seamless execution of secure quantized inference.}

\begin{figure}[h]
    \centering
    \includegraphics[scale=0.4]{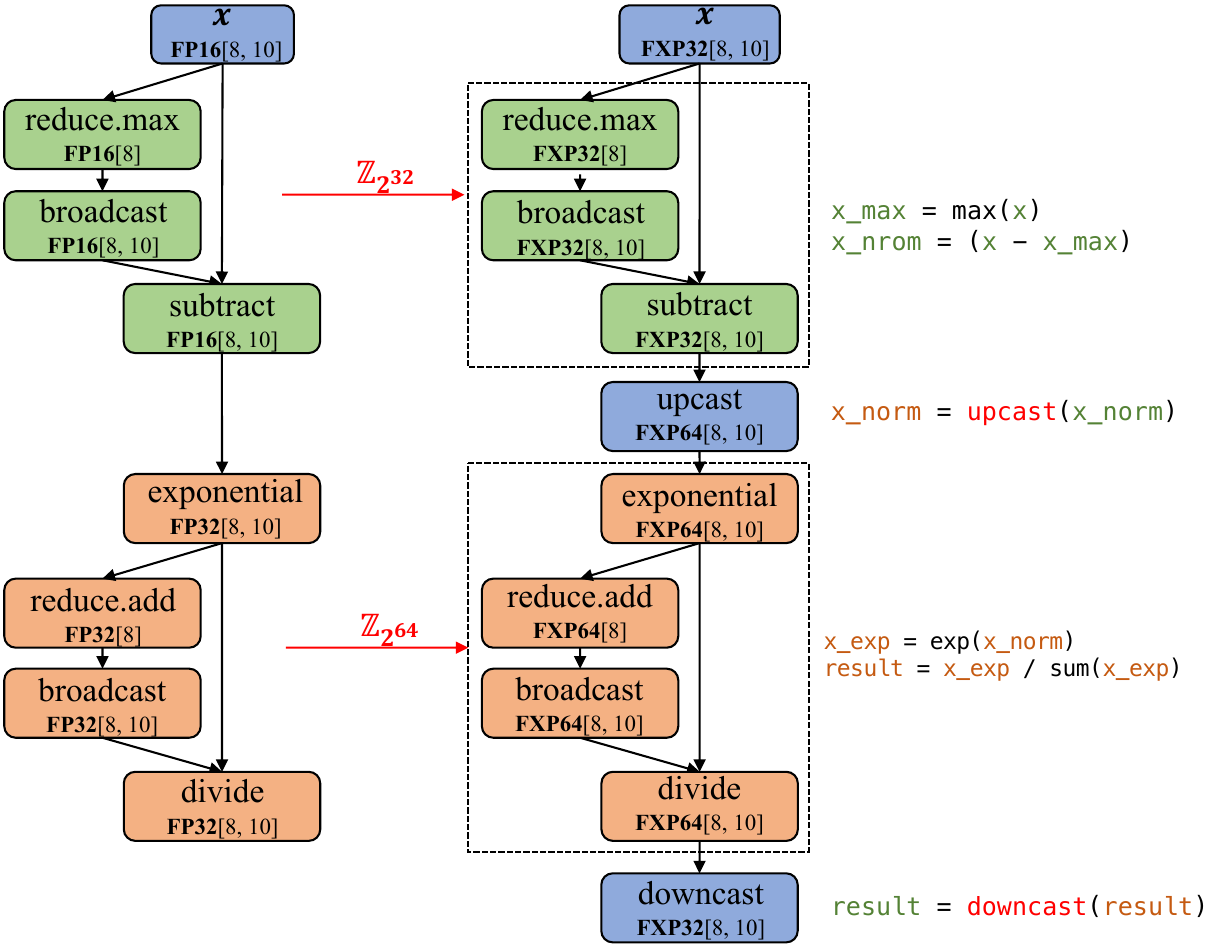}
    \caption{DAG of Softmax ($x \in \mathbb{R}^{8\times 10}$). Low-bit and high-bit computations are marked in Green and Orange, resp. }
    \label{fig:dag}
\end{figure}

Figure~\ref{fig:dag} depicts the DAG of the Softmax function, where the right part illustrates the modified quantization-aware variant in the domain of secure computation.
\camera{The \textbf{upcast} and \textbf{downcast} operations are automatically called to align types (i.e., different quantizations and rings) upon MPC.}

\hqcom{
\noindent\textbf{Dynamic Ring Support.}
In many MPC frameworks, including SPU, secure computations are limited to a \textbf{uniform, statically-configured} ring, a constraint that hampers the versatile deployment of fixed-point numbers with variable bitwidths.
To overcome this limitation, we devise a dynamic mapping strategy that maps plaintext data types from front-end to corresponding ciphertext types and, consequently the back-end rings of different sizes.
For instance, \texttt{FP16} and \texttt{FP32} are associated with \texttt{FXP32} over $\mathbb{Z}_{2^{32}}$ and \texttt{FXP64} over $\mathbb{Z}_{2^{64}}$, respectively.
As a result, the secure computations can be dynamically dispatched to the appropriate rings. 
}

\hqcom{
\noindent\textbf{Automatic Type Conversion.}
We further refine the compiler to track the input and compute types associated with each operator. As inference proceeds, variables are encoded using different types (i.e., different precisions and bitwidths). 
We exploit the MPC primitives to perform the interleaving type conversions that automatically maintain type consistency. For instance, if an operator (such as exponential) is specified with an input type of $\texttt{FXP32}$ and a compute type of $\texttt{FXP64}$, an upcast operation will seamlessly transition the input to $\texttt{FXP64}$ before carrying out further computations.
}

\section{Experiments}\label{sec:experiments}
We evaluate \name mainly from three aspects: 1) model utility (Section~\ref{sec:exp-utility}); 2) inference efficiency (Section~\ref{sec:exp-efficiency}); 3) scalability experiments and ablation studies (Section~\ref{sec:exp-extensive}).

\textbf{Experimental setup.}
We implement \name upon the framework \spu~\footnote{SPU: \url{https://github.com/secretflow/spu}.} that supports privacy-preserving machine learning. We use pure fixed-point arithmetic during the quantization and distillation procedure. 
We conduct the experiments on one CentOS 8 machine equipped with one AMD Ryzen CPU (32 cores and 3.60GHz) and 256GB of RAM.
We consider two network environments: 1) LAN setting with a bandwidth of 5Gbps and 0.4ms round-trip time; 2) WAN setting with a bandwidth of 400Mbps and 40ms round-trip time. 
We simulate the network environments using the Linux \textit{tc} tool.


\textbf{Model architectures and datasets.}
We use the pre-trained Bert models and GPT models in Hugging Face~\cite{huggingface}. 
For Bert, we use Bert-base
and Bert-large
pre-trained over BookCorpus~\cite{bookcorpus} and English Wikipedia~\cite{wikipedia} datasets. For GPT, we use GPT2-base
and GPT2-medium
pre-trained over the Wikitext-103 dataset~\cite{merity2016pointer}.
We evaluate Bert over RTE, CoLA, QQP and QNLI from GLUE benchmarks~\cite{GLUE}, and GPT2 on the validation set of Wikitext-103.
The detailed hyper-parameters for fine-tuning and distillation are in Appendix~\ref{append:hyperparameter}.

\textbf{Baselines.} 
We adopt secure inference upon SPU as the \textbf{vanilla baseline}.
The ablation models are denoted as $\name_{w/o\{a\}}$ with quantization, and $\name$ with both quantization and non-linear function approximation.
To make a more comprehensive comparison, we compare with two state-of-the-art work MPCFormer~\cite{li2023mpcformer} and PUMA~\cite{puma-2023}, which are similar to our setting.




\subsection{Utility Evaluation}\label{sec:exp-utility}
The evaluation of model utility is based on various accuracy metrics for downstream tasks. Concretely, we adopt Accuracy for RTE and QNLI, Matthews correlation for CoLA, F1 score for QQP, and Perplexity for Wikitext-103. 
In the GLUE benchmark, the input sequence length is set to 128 for Bert-base and Bert-large. For Wikitext-103, the input sequence length is set to 50 for GPT2-base and GPT2-medium.
Regarding MPCFormer, we explore two variants: Quad-alone and Quad+2ReLU.
For PUMA, the GeLU function is computed using their Poly approximation (cf., Appendix~\ref{append:poly-gelu-in-puma}).

The results are provided in Table~\ref{tab:utility}.
In general, with mere Quad approximation, \name (Quad) achieves similar results to that of MPCFormer (Quad) and slightly lower than the baseline without any quantization or approximation. 
The utility degradation is negligible on most datasets, except CoLA. The lower utility of both MPCFormer and \name on CoLA could be attributed to its smaller size, leading to unstable distillation performance. 
Notably we observe that with 2ReLU approximation, both MPCFormer and \name incur noticeable utility drops in Bert tasks. This is in line with the results reported in MPCFormer, thus indicating that Softmax is more sensitive to precision.
Regarding PUMA, it is worth noting that it incurs almost no accuracy drop due to the usage of a more accurate Polynomial approximation. However, as demonstrated in the following experiment, this improved accuracy comes at the cost of more communication overhead.
To balance between utility and efficiency, we mainly use Quad approximation for GeLU in \name.


\begin{table*}[h]
    \centering
    \setlength\tabcolsep{12pt}
    \caption{Model utility on GLUE benchmark for Bert and on Wikitext-103 dataset for GPT2. 
    }
    \scalebox{0.65}{
    \begin{tabular}{@{}c|c|cccc|cccc|cc@{}}
\toprule
\multirow{2}{*}{Method}    & \multirow{2}{*}{Approx.} & \multicolumn{4}{c|}{Bert-base}                                       & \multicolumn{4}{c|}{Bert-large}                                      & \multicolumn{1}{c|}{GPT2-base}      & GPT2-medium    \\ \cmidrule(l){3-12} 
                           &                          & RTE            & CoLA           & QQP            & QNLI ($\uparrow$) & RTE            & CoLA           & QQP            & QNLI ($\uparrow$) & \multicolumn{2}{c}{Wikitext-103 ($\downarrow$)}      \\ \midrule
Baseline                   & -                        & 68.59          & 57.06          & 87.96          & 91.62             & 72.56          & 63.09          & 88.52          & 92.58             & \multicolumn{1}{c|}{12.25}          & 10.60          \\ \midrule
\multirow{2}{*}{MPCFormer} & Quad                     & 67.85          & 54.47          & 87.76          & 91.68             & 71.86          & 57.53          & 88.34          & 92.53             & \multicolumn{1}{c|}{-}              & -              \\
                           & Quad+2ReLU               & 64.30          & 52.75          & 86.95          & 90.76             & 70.29          & 55.53          & 87.64          & 91.85             & \multicolumn{1}{c|}{-}              & -              \\ \midrule
PUMA                       & Poly                     & 68.47          & 56.96          & 87.95          & 91.48             & 72.56          & 62.60          & 88.50          & 92.55             & \multicolumn{1}{c|}{12.25}          & 10.49          \\ \midrule
$\name_{w/o\{a\}}$         & -                        & \textbf{67.87} & 54.17          & 87.15          & 91.74             & \textbf{72.55} & 56.25          & 88.22          & 92.58             & \multicolumn{1}{c|}{\textbf{12.99}} & \textbf{10.61} \\
\multirow{2}{*}{$\name$}   & Quad                     & 67.82          & \textbf{54.52} & \textbf{87.72} & \textbf{91.78}    & 71.84          & \textbf{56.45} & \textbf{88.23} & \textbf{92.58}    & \multicolumn{1}{c|}{13.78}          & 11.35          \\
                           & Quad+2ReLU               & 63.89          & 52.78          & 86.92          & 87.71             & 71.48          & 51.69          & 87.51          & 87.53             & \multicolumn{1}{c|}{-}              & -              \\ \bottomrule
\end{tabular}
}
    \label{tab:utility}
\end{table*}


    

\subsection{Efficiency Evaluation}\label{sec:exp-efficiency}
We measure the end-to-end runtime and concrete communication size to evaluate the efficiency. 
The experiments are conducted with a batch size of 1 in the LAN setting (\hqcom{refer to Appendix~\ref{append:supplementary-exps} for experiments in WAN}).
For Bert models, the input sequence length is set to 128. We choose the CoLA task as a representative since other tasks share the same model architecture as CoLA, resulting in similar inference overhead.
As for GPT2 models, we generate 1 new token with an input length of 32.
We run experiments for MPCFormer~\footnote{MPCFormer is configured to run on CPU for fair comparisons.} and \name with and without the Quad approximation of GeLU for a comprehensive comparison.

\begin{figure*}[ht!]
    \centering
    \includegraphics[width=0.76\linewidth]{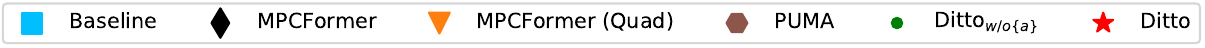}
    \subcaptionbox{Bert-base
    \label{fig:bert-base-efficiency}}[0.24\linewidth]
    {
        \includegraphics[width=1\linewidth]{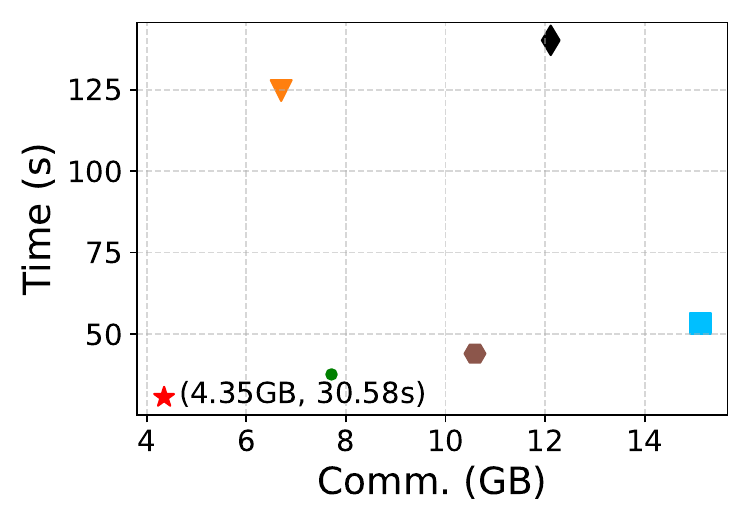}
    }
    \subcaptionbox{Bert-large
    \label{fig:bert-large-efficiency}}[0.24\linewidth]
    {
        \includegraphics[width=1\linewidth]{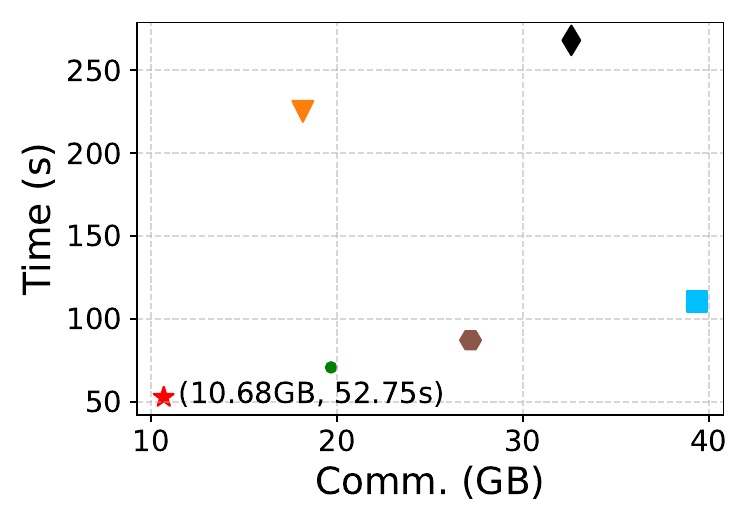}
    }
    \subcaptionbox{GPT2-base
    \label{fig:GPT2-base-efficiency}}[0.24\linewidth]
    {
        \includegraphics[width=1\linewidth]{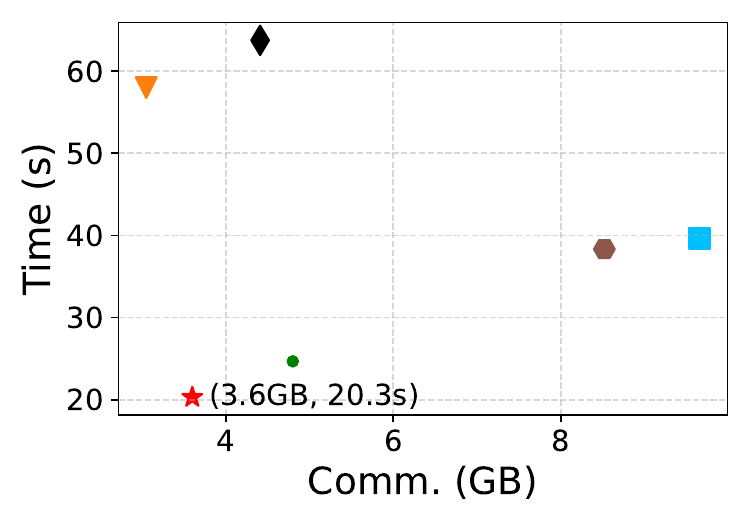}
    }
    \subcaptionbox{GPT2-medium
    \label{fig:GPT2-medium-efficiency}}[0.24\linewidth]
    {
        \includegraphics[width=1\linewidth]{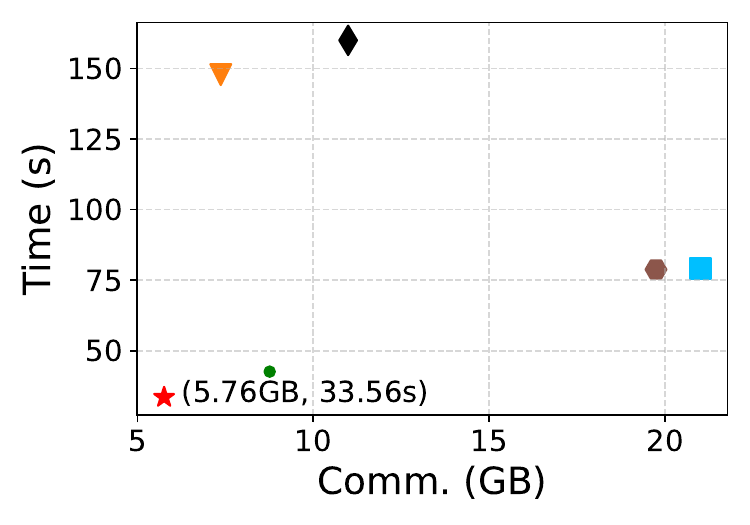}
    }
    \caption{Efficiency evaluation on Bert and GPT2 models. The closer to the bottom left corner, the better the performance.
    }
    \label{fig:efficiency}
\end{figure*}

\hqcom{In Figure~\ref{fig:efficiency}, the $x$- and $y$- axes represent communication size (\textbf{in GB}) and runtime (\textbf{in seconds}), respectively. The closer the dot is to the \textbf{bottom left} corner, the \textbf{better} the performance.}
On the four models, \name (marked in \textbf{red star}) generally has the best efficiency. 
Regarding communication size, $\name_{w/o\{a\}}$ incurs $1.37\sim 2.25\times$ lower communication than PUMA, and $1.25\sim 1.66\times$ than MPCFormer. 
When combined with Quad approximation, both MPCFormer (Quad) and \name incur lower communication than PUMA. Concretely, the communication size of \name is $1.28\sim 1.70\times$ lower than MPCFormer (Quad) and $2.37\sim 3.43\times$ than PUMA. 
Owing to the reduction of communication size, \name is $3.14\sim 4.40\times$ and $1.44\sim 2.35\times$ faster than MPCFormer (Quad) and PUMA, respectively. 
One exception is that MPCFormer incurs lower communication size on GPT2-base. This may be because GPT2 models have a larger vocabulary size, thus leading to a much higher overhead in the embedding layer for \name~\footnote{We convert the input token ids to one-hot vectors using MPC, while MPCFormer performs the conversion in the client locally.}.


\subsection{Extensive Experiments}\label{sec:exp-extensive}
We hereby evaluate the efficiency with varying input sequence length. The experiments for varying batch size and network environments, \hqcom{along with comparison to state-of-the-art 2PC works} are presented in Appendix~\ref{append:supplementary-exps}.

\begin{table}[h]
    \centering
    \tabcolsep=0.7mm
    \caption{Inference efficiency on Bert-base and GPT2-base with varying input sequence length. }\label{tab:efficiency-vary-input}
    \scalebox{0.6}{
    \begin{tabular}{@{}c|c|cccccccc@{}}
\toprule
\multirow{3}{*}{Model}     & \multirow{3}{*}{Method}   & \multicolumn{8}{c}{\#Input Length}                                                                                                                                                                \\ \cmidrule(l){3-10} 
                           &                          & \multicolumn{2}{c|}{32}                             & \multicolumn{2}{c|}{64}                             & \multicolumn{2}{c|}{128}                            & \multicolumn{2}{c}{256}         \\ \cmidrule(l){3-10} 
                           &                          & Comm.         & \multicolumn{1}{c|}{Time}           & Comm.         & \multicolumn{1}{c|}{Time}           & Comm.         & \multicolumn{1}{c|}{Time}           & Comm.          & Time           \\ \midrule
\multirow{5}{*}{Bert-base} & Baseline                 & 2.79          & \multicolumn{1}{c|}{13.57}          & 6.24          & \multicolumn{1}{c|}{27.32}          & 15.12         & \multicolumn{1}{c|}{53.17}          & 40.65          & 114.83         \\ \cmidrule(l){2-10} 
                           & MPCFormer                & 2.08          & \multicolumn{1}{c|}{35.15}          & 3.02          & \multicolumn{1}{c|}{63.94}          & 6.70          & \multicolumn{1}{c|}{124.81}         & 19.12          & 253.20         \\ \cmidrule(l){2-10} 
                           & PUMA                     & 2.16          & \multicolumn{1}{c|}{12.97}          & 4.65          & \multicolumn{1}{c|}{22.59}          & 10.59         & \multicolumn{1}{c|}{43.98}          & 26.07          & 88.70          \\ \cmidrule(l){2-10} 
                           & \multirow{2}{*}{$\name$} & \textbf{0.72} & \multicolumn{1}{c|}{\textbf{7.36}}  & \textbf{1.68} & \multicolumn{1}{c|}{\textbf{14.80}} & \textbf{4.35} & \multicolumn{1}{c|}{\textbf{30.58}} & \textbf{12.78} & \textbf{62.48} \\ \cmidrule(l){3-10} 
                           &                          & 3.00$\times$  & \multicolumn{1}{c|}{1.76$\times$}   & 2.77$\times$  & \multicolumn{1}{c|}{1.53$\times$}   & 2.43$\times$  & \multicolumn{1}{c|}{1.44$\times$}   & 2.04$\times$   & 1.42$\times$   \\ \midrule
\multirow{5}{*}{GPT2-base} & Baseline                 & 8.88          & \multicolumn{1}{c|}{37.88}          & 12.50         & \multicolumn{1}{c|}{52.63}          & 22.03         & \multicolumn{1}{c|}{84.04}          & 47.91          & 157.47         \\ \cmidrule(l){2-10} 
                           & MPCFormer                & \textbf{3.05} & \multicolumn{1}{c|}{57.57}          & \textbf{4.01} & \multicolumn{1}{c|}{104.97}         & \textbf{7.73} & \multicolumn{1}{c|}{182.20}         & 20.24          & 403.85         \\ \cmidrule(l){2-10} 
                           & PUMA                     & 8.61          & \multicolumn{1}{c|}{35.22}          & 11.95         & \multicolumn{1}{c|}{48.33}          & 17.57         & \multicolumn{1}{c|}{73.77}          & 33.00          & 131.33         \\ \cmidrule(l){2-10} 
                           & \multirow{2}{*}{$\name$} & 3.59          & \multicolumn{1}{c|}{\textbf{19.52}} & 5.18          & \multicolumn{1}{c|}{\textbf{29.41}} & 8.25          & \multicolumn{1}{c|}{\textbf{49.93}} & \textbf{17.44} & \textbf{93.52} \\ \cmidrule(l){3-10} 
                           &                          & 2.40$\times$  & \multicolumn{1}{c|}{1.80$\times$}   & 2.31$\times$  & \multicolumn{1}{c|}{1.64$\times$}   & 2.13$\times$  & \multicolumn{1}{c|}{1.48$\times$}   & 1.89$\times$   & 1.40$\times$   \\ \bottomrule
\end{tabular}
}
\end{table}

\textbf{Varying Input Sequence Length.}
The language models typically have conversation sentences as inputs, thus having different input lengths. We hereby conduct experiments with input length $\in \{32, 64, 128, 256\}$ on Bert-base and GPT2-base to make a more comprehensive evaluation.

The results are shown in Table~\ref{tab:efficiency-vary-input} (the speedup numbers are against PUMA). In general, the communication size of \name is about $2\sim 3\times$ lower than the state-of-the-art PUMA. Owing to the communication reduction, \name achieves a speedup of about $1.4\sim 1.8\times$ against PUMA and a speedup of about $2.9\sim 4.8\times$ against MPCFormer.

\textbf{Ablation Studies.}
We study the effects of quantization and approximation on Bert models. 
As shown in Table~\ref{tab:softmax-approx}, the quantization with the Quad approximation for GeLU generally results in negligible degradation in utility. 
The speedup achieved against the vanilla baseline is approximately $1.41\sim 1.56\times$ with quantization alone and $1.74\sim 2.09\times$ with the additional GeLU approximation.

\begin{table}[h]
    \centering
    \tabcolsep=0.7mm
    \caption{Ablation studies of \name on Bert models.}
    \scalebox{0.6}{
    \begin{tabular}{@{}c|c|ccccc|ccccc@{}}
\toprule
\multirow{2}{*}{Method}  & \multirow{2}{*}{Approx.} & \multicolumn{5}{c|}{Bert-base}                                                       & \multicolumn{5}{c}{Bert-large}                                                       \\ \cmidrule(l){3-12} 
                         &                          & RTE            & CoLA           & QQP            & QNLI           & \textit{Speedup} & RTE            & CoLA           & QQP            & QNLI           & \textit{Speedup} \\ \midrule
Baseline                  & -                        & 68.59          & 57.06          & 87.96          & 91.62          & -                & 72.56          & 63.09          & 88.52          & 92.58          & -                \\ \midrule
$\name_{w/o\{a\}}$       & -                        & 67.87          & 54.17          & 87.15          & 91.74          & 1.41$\times$     & 72.55          & 56.25          & 88.22          & 92.58          & 1.56$\times$     \\ \midrule
$\name$ & Quad                     & 67.82 & 54.52 & 87.72 & 91.78 & 1.74$\times$     & 71.84 & 56.45 & 88.23 & 92.57 & 2.09$\times$     \\ \bottomrule
\end{tabular}
}
    \label{tab:softmax-approx}
\end{table}


\section{Conclusion}\label{sec:conclusion}
In this paper, we propose a framework \name to enable secure quantization-aware inference of Transformer models. By incorporating MPC-friendly ML quantization and quantization-aware MPC execution, \name reduces the overhead and enhances the inference efficiency. In the future, we plan to investigate adopting more aggressive quantization methods, i.e., using lower bits in secure inference.
\section*{Acknowledgement}
We would like to thank the anonymous reviewers and program chairs for their invaluable comments. We would also like to acknowledge the insightful feedback and efforts provided by Ye Dong on the final version of our paper. We extend our appreciation to Wen-jie Lu, Zhicong Huang, and Cheng Hong for their significant contributions to SPU and Ditto. Lastly, we thank all members of the SecretFlow team for their supports throughout this project.
\section*{Impact Statements}
This paper presents work whose goal is to advance the field of Machine Learning. Our work aims to address the critical privacy concerns arising in the widely used Transformer-based inference. We employ cryptographic secure multi-party computation techniques, which are of provable security, to protect both the user query and model weights during the Transformer inference. However, it is also worth acknowledging the efficiency decay caused by secure computation. One notable contribution of our work is the novel optimizations that incorporate plaintext quantization and secure computations over dynamic rings. The advancements could have broader implications to make secure Transformer inference more practical in the real world.


\bibliography{icml24}

\begin{thebibliography}{50}
\providecommand{\natexlab}[1]{#1}
\providecommand{\url}[1]{\texttt{#1}}
\expandafter\ifx\csname urlstyle\endcsname\relax
  \providecommand{\doi}[1]{doi: #1}\else
  \providecommand{\doi}{doi: \begingroup \urlstyle{rm}\Url}\fi

\bibitem[Agrawal et~al.(2019)Agrawal, Shamsabadi, Kusner, and Gasc{\'{o}}n]{quotient}
Agrawal, N., Shamsabadi, A.~S., Kusner, M.~J., and Gasc{\'{o}}n, A.
\newblock {QUOTIENT:} two-party secure neural network training and prediction.
\newblock In Cavallaro, L., Kinder, J., Wang, X., and Katz, J. (eds.), \emph{Proceedings of the 2019 {ACM} {SIGSAC} Conference on Computer and Communications Security, {CCS} 2019, London, UK, November 11-15, 2019}, pp.\  1231--1247. {ACM}, 2019.
\newblock \doi{10.1145/3319535.3339819}.
\newblock URL \url{https://doi.org/10.1145/3319535.3339819}.

\bibitem[Akimoto et~al.(2023)Akimoto, Fukuchi, Akimoto, and Sakuma]{privformer}
Akimoto, Y., Fukuchi, K., Akimoto, Y., and Sakuma, J.
\newblock Privformer: Privacy-preserving transformer with mpc.
\newblock In \emph{2023 IEEE 8th European Symposium on Security and Privacy (EuroSP)}, pp.\  392--410, Los Alamitos, CA, USA, 2023. IEEE Computer Society.
\newblock \doi{10.1109/EuroSP57164.2023.00031}.
\newblock URL \url{https://doi.ieeecomputersociety.org/10.1109/EuroSP57164.2023.00031}.

\bibitem[Araki et~al.(2016)Araki, Furukawa, Lindell, Nof, and Ohara]{rss-16}
Araki, T., Furukawa, J., Lindell, Y., Nof, A., and Ohara, K.
\newblock High-throughput semi-honest secure three-party computation with an honest majority.
\newblock In Weippl, E.~R., Katzenbeisser, S., Kruegel, C., Myers, A.~C., and Halevi, S. (eds.), \emph{Proceedings of the 2016 {ACM} {SIGSAC} Conference on Computer and Communications Security, Vienna, Austria, October 24-28, 2016}, pp.\  805--817. {ACM}, 2016.
\newblock \doi{10.1145/2976749.2978331}.
\newblock URL \url{https://doi.org/10.1145/2976749.2978331}.

\bibitem[Bombari et~al.(2022)Bombari, Amani, and Mondelli]{over-parameterization-nips22}
Bombari, S., Amani, M.~H., and Mondelli, M.
\newblock Memorization and optimization in deep neural networks with minimum over-parameterization.
\newblock In \emph{NeurIPS}, 2022.
\newblock URL \url{http://papers.nips.cc/paper\_files/paper/2022/hash/323746f0ae2fbd8b6f500dc2d5c5f898-Abstract-Conference.html}.

\bibitem[Brown et~al.(2020)Brown, Mann, Ryder, Subbiah, Kaplan, Dhariwal, Neelakantan, Shyam, Sastry, Askell, Agarwal, Herbert-Voss, Krueger, Henighan, Child, Ramesh, Ziegler, Wu, Winter, Hesse, Chen, Sigler, Litwin, Gray, Chess, Clark, Berner, McCandlish, Radford, Sutskever, and Amodei]{chatgpt}
Brown, T.~B., Mann, B., Ryder, N., Subbiah, M., Kaplan, J., Dhariwal, P., Neelakantan, A., Shyam, P., Sastry, G., Askell, A., Agarwal, S., Herbert-Voss, A., Krueger, G., Henighan, T., Child, R., Ramesh, A., Ziegler, D.~M., Wu, J., Winter, C., Hesse, C., Chen, M., Sigler, E., Litwin, M., Gray, S., Chess, B., Clark, J., Berner, C., McCandlish, S., Radford, A., Sutskever, I., and Amodei, D.
\newblock Language models are few-shot learners, 2020.

\bibitem[Byali et~al.(2020)Byali, Chaudhari, Patra, and Suresh]{flash-20}
Byali, M., Chaudhari, H., Patra, A., and Suresh, A.
\newblock {FLASH:} fast and robust framework for privacy-preserving machine learning.
\newblock \emph{Proc. Priv. Enhancing Technol.}, 2020\penalty0 (2):\penalty0 459--480, 2020.
\newblock \doi{10.2478/popets-2020-0036}.
\newblock URL \url{https://doi.org/10.2478/popets-2020-0036}.

\bibitem[Catrina \& Saxena(2010)Catrina and Saxena]{catrina-10}
Catrina, O. and Saxena, A.
\newblock Secure computation with fixed-point numbers.
\newblock In Sion, R. (ed.), \emph{Financial Cryptography and Data Security, 14th International Conference, {FC} 2010, Tenerife, Canary Islands, Spain, January 25-28, 2010, Revised Selected Papers}, volume 6052 of \emph{Lecture Notes in Computer Science}, pp.\  35--50. Springer, 2010.
\newblock \doi{10.1007/978-3-642-14577-3\_6}.
\newblock URL \url{https://doi.org/10.1007/978-3-642-14577-3\_6}.

\bibitem[Chen et~al.(2021)Chen, Xie, and He]{slef-supervised-vit-ICCV21}
Chen, X., Xie, S., and He, K.
\newblock An empirical study of training self-supervised vision transformers.
\newblock In \emph{2021 {IEEE/CVF} International Conference on Computer Vision, {ICCV} 2021, Montreal, QC, Canada, October 10-17, 2021}, pp.\  9620--9629. {IEEE}, 2021.
\newblock \doi{10.1109/ICCV48922.2021.00950}.
\newblock URL \url{https://doi.org/10.1109/ICCV48922.2021.00950}.

\bibitem[Chou et~al.(2018)Chou, Beal, Levy, Yeung, Haque, and Fei{-}Fei]{faster-CryptoNets}
Chou, E., Beal, J., Levy, D., Yeung, S., Haque, A., and Fei{-}Fei, L.
\newblock Faster cryptonets: Leveraging sparsity for real-world encrypted inference.
\newblock \emph{CoRR}, abs/1811.09953, 2018.
\newblock URL \url{http://arxiv.org/abs/1811.09953}.

\bibitem[Dalskov et~al.(2020)Dalskov, Escudero, and Keller]{secureq8}
Dalskov, A. P.~K., Escudero, D., and Keller, M.
\newblock Secure evaluation of quantized neural networks.
\newblock \emph{Proc. Priv. Enhancing Technol.}, 2020\penalty0 (4):\penalty0 355--375, 2020.
\newblock \doi{10.2478/POPETS-2020-0077}.
\newblock URL \url{https://doi.org/10.2478/popets-2020-0077}.

\bibitem[Dalskov et~al.(2021)Dalskov, Escudero, and Keller]{fantastic-four-21}
Dalskov, A. P.~K., Escudero, D., and Keller, M.
\newblock Fantastic four: Honest-majority four-party secure computation with malicious security.
\newblock In Bailey, M. and Greenstadt, R. (eds.), \emph{30th {USENIX} Security Symposium, {USENIX} Security 2021, August 11-13, 2021}, pp.\  2183--2200. {USENIX} Association, 2021.
\newblock URL \url{https://www.usenix.org/conference/usenixsecurity21/presentation/dalskov}.

\bibitem[Dettmers et~al.(2022)Dettmers, Lewis, Belkada, and Zettlemoyer]{llm.int8-22}
Dettmers, T., Lewis, M., Belkada, Y., and Zettlemoyer, L.
\newblock Llm.int8(): 8-bit matrix multiplication for transformers at scale.
\newblock \emph{CoRR}, abs/2208.07339, 2022.
\newblock \doi{10.48550/arXiv.2208.07339}.
\newblock URL \url{https://doi.org/10.48550/arXiv.2208.07339}.

\bibitem[Devlin et~al.(2019)Devlin, Chang, Lee, and Toutanova]{bert-19}
Devlin, J., Chang, M., Lee, K., and Toutanova, K.
\newblock {BERT:} pre-training of deep bidirectional transformers for language understanding.
\newblock In Burstein, J., Doran, C., and Solorio, T. (eds.), \emph{Proceedings of the 2019 Conference of the North American Chapter of the Association for Computational Linguistics: Human Language Technologies, {NAACL-HLT} 2019, Minneapolis, MN, USA, June 2-7, 2019, Volume 1 (Long and Short Papers)}, pp.\  4171--4186. Association for Computational Linguistics, 2019.
\newblock \doi{10.18653/v1/n19-1423}.
\newblock URL \url{https://doi.org/10.18653/v1/n19-1423}.

\bibitem[Dong et~al.(2023)Dong, Lu, Zheng, Wu, Zhao, Tan, Huang, Hong, Wei, and Chen]{puma-2023}
Dong, Y., Lu, W., Zheng, Y., Wu, H., Zhao, D., Tan, J., Huang, Z., Hong, C., Wei, T., and Chen, W.
\newblock {PUMA:} secure inference of llama-7b in five minutes.
\newblock \emph{CoRR}, abs/2307.12533, 2023.
\newblock \doi{10.48550/arXiv.2307.12533}.
\newblock URL \url{https://doi.org/10.48550/arXiv.2307.12533}.

\bibitem[Dosovitskiy et~al.(2021)Dosovitskiy, Beyer, Kolesnikov, Weissenborn, Zhai, Unterthiner, Dehghani, Minderer, Heigold, Gelly, Uszkoreit, and Houlsby]{vit-21}
Dosovitskiy, A., Beyer, L., Kolesnikov, A., Weissenborn, D., Zhai, X., Unterthiner, T., Dehghani, M., Minderer, M., Heigold, G., Gelly, S., Uszkoreit, J., and Houlsby, N.
\newblock An image is worth 16x16 words: Transformers for image recognition at scale.
\newblock In \emph{9th International Conference on Learning Representations, {ICLR} 2021, Virtual Event, Austria, May 3-7, 2021}. OpenReview.net, 2021.
\newblock URL \url{https://openreview.net/forum?id=YicbFdNTTy}.

\bibitem[Escudero et~al.(2020)Escudero, Ghosh, Keller, Rachuri, and Scholl]{mixed-ab-circuit-20}
Escudero, D., Ghosh, S., Keller, M., Rachuri, R., and Scholl, P.
\newblock Improved primitives for {MPC} over mixed arithmetic-binary circuits.
\newblock In Micciancio, D. and Ristenpart, T. (eds.), \emph{Advances in Cryptology - {CRYPTO} 2020 - 40th Annual International Cryptology Conference, {CRYPTO} 2020, Santa Barbara, CA, USA, August 17-21, 2020, Proceedings, Part {II}}, volume 12171 of \emph{Lecture Notes in Computer Science}, pp.\  823--852. Springer, 2020.
\newblock \doi{10.1007/978-3-030-56880-1\_29}.
\newblock URL \url{https://doi.org/10.1007/978-3-030-56880-1\_29}.

\bibitem[Frantar et~al.(2022)Frantar, Ashkboos, Hoefler, and Alistarh]{GPTQ-22}
Frantar, E., Ashkboos, S., Hoefler, T., and Alistarh, D.
\newblock {GPTQ:} accurate post-training quantization for generative pre-trained transformers.
\newblock \emph{CoRR}, abs/2210.17323, 2022.
\newblock \doi{10.48550/arXiv.2210.17323}.
\newblock URL \url{https://doi.org/10.48550/arXiv.2210.17323}.

\bibitem[Frantar et~al.(2023)Frantar, Ashkboos, Hoefler, and Alistarh]{OPTQ-23}
Frantar, E., Ashkboos, S., Hoefler, T., and Alistarh, D.
\newblock {OPTQ:} accurate quantization for generative pre-trained transformers.
\newblock In \emph{The Eleventh International Conference on Learning Representations, {ICLR} 2023, Kigali, Rwanda, May 1-5, 2023}. OpenReview.net, 2023.
\newblock URL \url{https://openreview.net/pdf?id=tcbBPnfwxS}.

\bibitem[Gholami et~al.(2021)Gholami, Kim, Dong, Yao, Mahoney, and Keutzer]{quantization-survey-21}
Gholami, A., Kim, S., Dong, Z., Yao, Z., Mahoney, M.~W., and Keutzer, K.
\newblock A survey of quantization methods for efficient neural network inference.
\newblock \emph{CoRR}, abs/2103.13630, 2021.
\newblock URL \url{https://arxiv.org/abs/2103.13630}.

\bibitem[Hao et~al.(2022)Hao, Li, Chen, Xing, Xu, and Zhang]{hao2022iron}
Hao, M., Li, H., Chen, H., Xing, P., Xu, G., and Zhang, T.
\newblock Iron: Private inference on transformers.
\newblock In Oh, A.~H., Agarwal, A., Belgrave, D., and Cho, K. (eds.), \emph{Advances in Neural Information Processing Systems}, 2022.
\newblock URL \url{https://openreview.net/forum?id=deyqjpcTfsG}.

\bibitem[Huang et~al.(2022)Huang, Lu, Hong, and Ding]{cheetah-22}
Huang, Z., Lu, W., Hong, C., and Ding, J.
\newblock Cheetah: Lean and fast secure two-party deep neural network inference.
\newblock In Butler, K. R.~B. and Thomas, K. (eds.), \emph{31st {USENIX} Security Symposium, {USENIX} Security 2022, Boston, MA, USA, August 10-12, 2022}, pp.\  809--826. {USENIX} Association, 2022.
\newblock URL \url{https://www.usenix.org/conference/usenixsecurity22/presentation/huang-zhicong}.

\bibitem[Jiao et~al.(2020)Jiao, Yin, Shang, Jiang, Chen, Li, Wang, and Liu]{tinybert-emnlp20}
Jiao, X., Yin, Y., Shang, L., Jiang, X., Chen, X., Li, L., Wang, F., and Liu, Q.
\newblock Tinybert: Distilling {BERT} for natural language understanding.
\newblock In Cohn, T., He, Y., and Liu, Y. (eds.), \emph{Findings of the Association for Computational Linguistics: {EMNLP} 2020, Online Event, 16-20 November 2020}, volume {EMNLP} 2020 of \emph{Findings of {ACL}}, pp.\  4163--4174. Association for Computational Linguistics, 2020.
\newblock \doi{10.18653/v1/2020.findings-emnlp.372}.
\newblock URL \url{https://doi.org/10.18653/v1/2020.findings-emnlp.372}.

\bibitem[Kim et~al.(2021)Kim, Gholami, Yao, Mahoney, and Keutzer]{i-bert-icml21}
Kim, S., Gholami, A., Yao, Z., Mahoney, M.~W., and Keutzer, K.
\newblock {I-BERT:} integer-only {BERT} quantization.
\newblock In Meila, M. and Zhang, T. (eds.), \emph{Proceedings of the 38th International Conference on Machine Learning, {ICML} 2021, 18-24 July 2021, Virtual Event}, volume 139 of \emph{Proceedings of Machine Learning Research}, pp.\  5506--5518. {PMLR}, 2021.
\newblock URL \url{http://proceedings.mlr.press/v139/kim21d.html}.

\bibitem[Li et~al.(2023)Li, Wang, Shao, Guo, Xing, and Zhang]{li2023mpcformer}
Li, D., Wang, H., Shao, R., Guo, H., Xing, E., and Zhang, H.
\newblock {MPCFORMER}: {FAST}, {PERFORMANT} {AND} {PRIVATE} {TRANSFORMER} {INFERENCE} {WITH} {MPC}.
\newblock In \emph{The Eleventh International Conference on Learning Representations}, 2023.
\newblock URL \url{https://openreview.net/forum?id=CWmvjOEhgH-}.

\bibitem[Liang et~al.(2023)Liang, Wang, Zhang, Xing, Xu, and Zhang]{liang2023merge}
Liang, Z., Wang, P., Zhang, R., Xing, L., Xu, N., and Zhang, S.
\newblock Merge: Fast private text generation, 2023.

\bibitem[Liu \& Liu(2023)Liu and Liu]{liu2023llms}
Liu, X. and Liu, Z.
\newblock Llms can understand encrypted prompt: Towards privacy-computing friendly transformers, 2023.

\bibitem[Lu et~al.(2020)Lu, Fang, Huang, Hong, Chen, Qu, Zhou, and Ren]{rsqrt}
Lu, W., Fang, Y., Huang, Z., Hong, C., Chen, C., Qu, H., Zhou, Y., and Ren, K.
\newblock Faster secure multiparty computation of adaptive gradient descent.
\newblock In Zhang, B., Popa, R.~A., Zaharia, M., Gu, G., and Ji, S. (eds.), \emph{PPMLP'20: Proceedings of the 2020 Workshop on Privacy-Preserving Machine Learning in Practice, Virtual Event, USA, November, 2020}, pp.\  47--49. {ACM}, 2020.
\newblock \doi{10.1145/3411501.3419427}.
\newblock URL \url{https://doi.org/10.1145/3411501.3419427}.

\bibitem[Lu et~al.(2023)Lu, Huang, Gu, Li, Liu, Ren, Hong, Wei, and Chen]{bumblebee}
Lu, W., Huang, Z., Gu, Z., Li, J., Liu, J., Ren, K., Hong, C., Wei, T., and Chen, W.
\newblock Bumblebee: Secure two-party inference framework for large transformers.
\newblock \emph{{IACR} Cryptol. ePrint Arch.}, pp.\  1678, 2023.
\newblock URL \url{https://eprint.iacr.org/2023/1678}.

\bibitem[Ma et~al.(2023)Ma, Zheng, Feng, Zhao, Wu, Fang, Tan, Yu, Zhang, and Wang]{spu-atc23}
Ma, J., Zheng, Y., Feng, J., Zhao, D., Wu, H., Fang, W., Tan, J., Yu, C., Zhang, B., and Wang, L.
\newblock Secretflow-spu: {A} performant and user-friendly framework for privacy-preserving machine learning.
\newblock In Lawall, J. and Williams, D. (eds.), \emph{2023 {USENIX} Annual Technical Conference, {USENIX} {ATC} 2023, Boston, MA, USA, July 10-12, 2023}, pp.\  17--33. {USENIX} Association, 2023.
\newblock URL \url{https://www.usenix.org/conference/atc23/presentation/ma}.

\bibitem[Merity et~al.(2016)Merity, Xiong, Bradbury, and Socher]{merity2016pointer}
Merity, S., Xiong, C., Bradbury, J., and Socher, R.
\newblock Pointer sentinel mixture models, 2016.

\bibitem[Micikevicius et~al.(2017)Micikevicius, Narang, Alben, Diamos, Elsen, Garc{\'{\i}}a, Ginsburg, Houston, Kuchaiev, Venkatesh, and Wu]{mp-training-19}
Micikevicius, P., Narang, S., Alben, J., Diamos, G.~F., Elsen, E., Garc{\'{\i}}a, D., Ginsburg, B., Houston, M., Kuchaiev, O., Venkatesh, G., and Wu, H.
\newblock Mixed precision training.
\newblock \emph{CoRR}, abs/1710.03740, 2017.
\newblock URL \url{http://arxiv.org/abs/1710.03740}.

\bibitem[Mohassel \& Rindal(2018)Mohassel and Rindal]{aby3}
Mohassel, P. and Rindal, P.
\newblock Aby3: A mixed protocol framework for machine learning.
\newblock In \emph{Proceedings of the 2018 ACM SIGSAC Conference on Computer and Communications Security}, pp.\  35–52, New York, NY, USA, 2018. Association for Computing Machinery.
\newblock ISBN 9781450356930.
\newblock \doi{10.1145/3243734.3243760}.
\newblock URL \url{https://doi.org/10.1145/3243734.3243760}.

\bibitem[Mohassel \& Zhang(2017)Mohassel and Zhang]{secureml-17}
Mohassel, P. and Zhang, Y.
\newblock Secureml: {A} system for scalable privacy-preserving machine learning.
\newblock In \emph{2017 {IEEE} Symposium on Security and Privacy, {SP} 2017, San Jose, CA, USA, May 22-26, 2017}, pp.\  19--38. {IEEE} Computer Society, 2017.
\newblock \doi{10.1109/SP.2017.12}.
\newblock URL \url{https://doi.org/10.1109/SP.2017.12}.

\bibitem[Patra et~al.(2021)Patra, Schneider, Suresh, and Yalame]{aby2.0-21}
Patra, A., Schneider, T., Suresh, A., and Yalame, H.
\newblock {ABY2.0:} improved mixed-protocol secure two-party computation.
\newblock In Bailey, M. and Greenstadt, R. (eds.), \emph{30th {USENIX} Security Symposium, {USENIX} Security 2021, August 11-13, 2021}, pp.\  2165--2182. {USENIX} Association, 2021.
\newblock URL \url{https://www.usenix.org/conference/usenixsecurity21/presentation/patra}.

\bibitem[Radford et~al.(2019)Radford, Wu, Child, Luan, Amodei, and Sutskever]{gpt2-19}
Radford, A., Wu, J., Child, R., Luan, D., Amodei, D., and Sutskever, I.
\newblock Language models are unsupervised multitask learners.
\newblock 2019.

\bibitem[Riazi et~al.(2019)Riazi, Samragh, Chen, Laine, Lauter, and Koushanfar]{XONN}
Riazi, M.~S., Samragh, M., Chen, H., Laine, K., Lauter, K.~E., and Koushanfar, F.
\newblock {XONN:} xnor-based oblivious deep neural network inference.
\newblock In Heninger, N. and Traynor, P. (eds.), \emph{28th {USENIX} Security Symposium, {USENIX} Security 2019, Santa Clara, CA, USA, August 14-16, 2019}, pp.\  1501--1518. {USENIX} Association, 2019.
\newblock URL \url{https://www.usenix.org/conference/usenixsecurity19/presentation/riazi}.

\bibitem[Shamir(1979)]{shamir-79}
Shamir, A.
\newblock How to share a secret.
\newblock \emph{Commun. {ACM}}, 22\penalty0 (11):\penalty0 612--613, 1979.
\newblock \doi{10.1145/359168.359176}.
\newblock URL \url{https://doi.org/10.1145/359168.359176}.

\bibitem[Shokri et~al.(2017)Shokri, Stronati, Song, and Shmatikov]{membership-inference-17}
Shokri, R., Stronati, M., Song, C., and Shmatikov, V.
\newblock Membership inference attacks against machine learning models.
\newblock In \emph{2017 {IEEE} Symposium on Security and Privacy, {SP} 2017, San Jose, CA, USA, May 22-26, 2017}, pp.\  3--18. {IEEE} Computer Society, 2017.
\newblock \doi{10.1109/SP.2017.41}.
\newblock URL \url{https://doi.org/10.1109/SP.2017.41}.

\bibitem[Tan et~al.(2021)Tan, Knott, Tian, and Wu]{cryptgpu-sp21}
Tan, S., Knott, B., Tian, Y., and Wu, D.~J.
\newblock Cryptgpu: Fast privacy-preserving machine learning on the {GPU}.
\newblock In \emph{42nd {IEEE} Symposium on Security and Privacy, {SP} 2021, San Francisco, CA, USA, 24-27 May 2021}, pp.\  1021--1038. {IEEE}, 2021.
\newblock \doi{10.1109/SP40001.2021.00098}.
\newblock URL \url{https://doi.org/10.1109/SP40001.2021.00098}.

\bibitem[Vaswani et~al.(2017)Vaswani, Shazeer, Parmar, Uszkoreit, Jones, Gomez, Kaiser, and Polosukhin]{attention-all-need-17}
Vaswani, A., Shazeer, N., Parmar, N., Uszkoreit, J., Jones, L., Gomez, A.~N., Kaiser, L., and Polosukhin, I.
\newblock Attention is all you need.
\newblock In Guyon, I., von Luxburg, U., Bengio, S., Wallach, H.~M., Fergus, R., Vishwanathan, S. V.~N., and Garnett, R. (eds.), \emph{Advances in Neural Information Processing Systems 30: Annual Conference on Neural Information Processing Systems 2017, December 4-9, 2017, Long Beach, CA, {USA}}, pp.\  5998--6008, 2017.
\newblock URL \url{https://proceedings.neurips.cc/paper/2017/hash/3f5ee243547dee91fbd053c1c4a845aa-Abstract.html}.

\bibitem[Wagh et~al.(2021)Wagh, Tople, Benhamouda, Kushilevitz, Mittal, and Rabin]{falcon-20}
Wagh, S., Tople, S., Benhamouda, F., Kushilevitz, E., Mittal, P., and Rabin, T.
\newblock Falcon: Honest-majority maliciously secure framework for private deep learning.
\newblock \emph{Proc. Priv. Enhancing Technol.}, 2021\penalty0 (1):\penalty0 188--208, 2021.
\newblock \doi{10.2478/popets-2021-0011}.
\newblock URL \url{https://doi.org/10.2478/popets-2021-0011}.

\bibitem[Wang et~al.(2019)Wang, Singh, Michael, Hill, Levy, and Bowman]{GLUE}
Wang, A., Singh, A., Michael, J., Hill, F., Levy, O., and Bowman, S.~R.
\newblock {GLUE:} {A} multi-task benchmark and analysis platform for natural language understanding.
\newblock In \emph{7th International Conference on Learning Representations, {ICLR} 2019, New Orleans, LA, USA, May 6-9, 2019}. OpenReview.net, 2019.
\newblock URL \url{https://openreview.net/forum?id=rJ4km2R5t7}.

\bibitem[{Wikipedia contributors}(2004)]{wikipedia}
{Wikipedia contributors}.
\newblock Plagiarism --- {W}ikipedia{,} the free encyclopedia, 2004.
\newblock URL \url{https://en.wikipedia.org/w/index.php?title=Plagiarism&oldid=5139350}.
\newblock [Online; accessed 22-July-2004].

\bibitem[Wolf et~al.(2020)Wolf, Debut, Sanh, Chaumond, Delangue, Moi, Cistac, Rault, Louf, Funtowicz, Davison, Shleifer, von Platen, Ma, Jernite, Plu, Xu, Scao, Gugger, Drame, Lhoest, and Rush]{huggingface}
Wolf, T., Debut, L., Sanh, V., Chaumond, J., Delangue, C., Moi, A., Cistac, P., Rault, T., Louf, R., Funtowicz, M., Davison, J., Shleifer, S., von Platen, P., Ma, C., Jernite, Y., Plu, J., Xu, C., Scao, T.~L., Gugger, S., Drame, M., Lhoest, Q., and Rush, A.~M.
\newblock Transformers: State-of-the-art natural language processing.
\newblock In \emph{Proceedings of the 2020 Conference on Empirical Methods in Natural Language Processing: System Demonstrations}, pp.\  38--45, Online, October 2020. Association for Computational Linguistics.
\newblock URL \url{https://www.aclweb.org/anthology/2020.emnlp-demos.6}.

\bibitem[Yao(1986)]{yao-mpc}
Yao, A.~C.
\newblock How to generate and exchange secrets (extended abstract).
\newblock In \emph{27th Annual Symposium on Foundations of Computer Science, Toronto, Canada, 27-29 October 1986}, pp.\  162--167. {IEEE} Computer Society, 1986.
\newblock \doi{10.1109/SFCS.1986.25}.
\newblock URL \url{https://doi.org/10.1109/SFCS.1986.25}.

\bibitem[Yao et~al.(2021)Yao, Dong, Zheng, Gholami, Yu, Tan, Wang, Huang, Wang, Mahoney, and Keutzer]{hawqv3-icml21}
Yao, Z., Dong, Z., Zheng, Z., Gholami, A., Yu, J., Tan, E., Wang, L., Huang, Q., Wang, Y., Mahoney, M.~W., and Keutzer, K.
\newblock {HAWQ-V3:} dyadic neural network quantization.
\newblock In Meila, M. and Zhang, T. (eds.), \emph{Proceedings of the 38th International Conference on Machine Learning, {ICML} 2021, 18-24 July 2021, Virtual Event}, volume 139 of \emph{Proceedings of Machine Learning Research}, pp.\  11875--11886. {PMLR}, 2021.
\newblock URL \url{http://proceedings.mlr.press/v139/yao21a.html}.

\bibitem[Yao et~al.(2022)Yao, Aminabadi, Zhang, Wu, Li, and He]{ZeroQuant-nips22}
Yao, Z., Aminabadi, R.~Y., Zhang, M., Wu, X., Li, C., and He, Y.
\newblock Zeroquant: Efficient and affordable post-training quantization for large-scale transformers.
\newblock In \emph{NeurIPS}, 2022.
\newblock URL \url{http://papers.nips.cc/paper\_files/paper/2022/hash/adf7fa39d65e2983d724ff7da57f00ac-Abstract-Conference.html}.

\bibitem[Zeng et~al.(2023)Zeng, Li, Xiong, Tong, Lu, Tan, Wang, and Huang]{mpcvit-23}
Zeng, W., Li, M., Xiong, W., Tong, T., Lu, W., Tan, J., Wang, R., and Huang, R.
\newblock Mpcvit: Searching for accurate and efficient mpc-friendly vision transformer with heterogeneous attention.
\newblock In \emph{{IEEE/CVF} International Conference on Computer Vision, {ICCV} 2023, Paris, France, October 1-6, 2023}, pp.\  5029--5040. {IEEE}, 2023.
\newblock \doi{10.1109/ICCV51070.2023.00466}.
\newblock URL \url{https://doi.org/10.1109/ICCV51070.2023.00466}.

\bibitem[Zhang et~al.(2023)Zhang, Chen, Kundu, Li, and Beerel]{SAL-ViT-23}
Zhang, Y., Chen, D., Kundu, S., Li, C., and Beerel, P.~A.
\newblock Sal-vit: Towards latency efficient private inference on vit using selective attention search with a learnable softmax approximation.
\newblock In \emph{{IEEE/CVF} International Conference on Computer Vision, {ICCV} 2023, Paris, France, October 1-6, 2023}, pp.\  5093--5102. {IEEE}, 2023.
\newblock \doi{10.1109/ICCV51070.2023.00472}.
\newblock URL \url{https://doi.org/10.1109/ICCV51070.2023.00472}.

\bibitem[Zhu et~al.(2015)Zhu, Kiros, Zemel, Salakhutdinov, Urtasun, Torralba, and Fidler]{bookcorpus}
Zhu, Y., Kiros, R., Zemel, R., Salakhutdinov, R., Urtasun, R., Torralba, A., and Fidler, S.
\newblock Aligning books and movies: Towards story-like visual explanations by watching movies and reading books.
\newblock In \emph{The IEEE International Conference on Computer Vision (ICCV)}, December 2015.

\end{thebibliography}
\bibliographystyle{icml2024}

\newpage
\appendix
\onecolumn
\appendix
\section{Supplementary Experiments}\label{append:supplementary-exps}
\paragraph{Varying Batch Size.}
We evaluate the secure inference with batch size $\in \{1, 2, 4, 8\}$ on Bert-base model.
As shown in Table~\ref{tab:efficiency-varying-batch-size}, the communication size and runtime increase about linearly to the batch size for all the methods. \name remains about $1.4\times$ and $4.0\times$ faster than PUMA and MPCFormer, respectively. 

\begin{table}[h]
    \centering
    \setlength\tabcolsep{6pt}
    \caption{Inference efficiency on Bert-base with varying batch size. The input length is set to 128.}
    \label{tab:efficiency-varying-batch-size}
    \scalebox{0.8}{
    \begin{tabular}{@{}c|c|cccccccc@{}}
\toprule
\multirow{3}{*}{Model}     & \multirow{3}{*}{Method}  & \multicolumn{8}{c}{\#Batch Size}                                                                                                                                                                     \\ \cmidrule(l){3-10} 
                           &                          & \multicolumn{2}{c|}{1}                              & \multicolumn{2}{c|}{2}                              & \multicolumn{2}{c|}{4}                                & \multicolumn{2}{c}{8}            \\ \cmidrule(l){3-10} 
                           &                          & Comm.         & \multicolumn{1}{c|}{Time}           & Comm.         & \multicolumn{1}{c|}{Time}           & Comm.          & \multicolumn{1}{c|}{Time}            & Comm.          & Time            \\ \midrule
\multirow{5}{*}{Bert-base} & Baseline                 & 15.12         & \multicolumn{1}{c|}{53.17}          & 30.21         & \multicolumn{1}{c|}{97.56}          & 60.49          & \multicolumn{1}{c|}{185.63}          & 120.77         & 365.76          \\ \cmidrule(l){2-10} 
                           & MPCFormer                & 6.70          & \multicolumn{1}{c|}{124.81}         & 11.77         & \multicolumn{1}{c|}{227.82}         & 21.93          & \multicolumn{1}{c|}{432.06}          & 42.25          & 839.83          \\ \cmidrule(l){2-10} 
                           & PUMA                     & 10.59         & \multicolumn{1}{c|}{43.98}          & 21.20         & \multicolumn{1}{c|}{79.70}          & 42.27          & \multicolumn{1}{c|}{153.88}          & 84.65          & 297.90          \\ \cmidrule(l){2-10} 
                           & \multirow{2}{*}{$\name$} & \textbf{4.35} & \multicolumn{1}{c|}{\textbf{30.58}} & \textbf{8.68} & \multicolumn{1}{c|}{\textbf{56.29}} & \textbf{17.35} & \multicolumn{1}{c|}{\textbf{107.52}} & \textbf{34.96} & \textbf{208.63} \\ \cmidrule(l){3-10} 
                           &                          & 2.43$\times$  & \multicolumn{1}{c|}{1.44$\times$}   & 2.44$\times$  & \multicolumn{1}{c|}{1.42$\times$}   & 2.44$\times$   & \multicolumn{1}{c|}{1.43$\times$}    & 2.42$\times$   & 1.43$\times$    \\ \bottomrule
\end{tabular}
}
\end{table}

\paragraph{Varying Network Environment.}
We evaluate the secure inference under two different network settings, i.e., LAN and WAN. Since secure inference based on MPC is communication-bound, the network status has a significant effect on the efficiency.
As shown in Table~\ref{tab:varying-network}, the runtime increases dramatically in WAN, which is nearly $10\times$ that in LAN. This is because WAN has a smaller bandwidth and higher latency. Compared to PUMA, \name is still $1.46\sim 1.53\times$ faster in WAN due to the reduction of communication overhead.

\begin{table}[h]
    \centering
    \setlength\tabcolsep{8pt}
    \caption{Inference efficiency of Bert-base and GPT2-base under different network environments.}
    \label{tab:varying-network}
    \scalebox{0.8}{
    \begin{tabular}{@{}c|c|ccccc@{}}
\toprule
\multirow{2}{*}{Network}   & \multirow{2}{*}{Model} & \multicolumn{5}{c}{Runtime (s)}                                                                                               \\ \cmidrule(l){3-7} 
                           &                        & \multicolumn{1}{c|}{Baseline} & \multicolumn{1}{c|}{MPCFormer} & \multicolumn{1}{c|}{PUMA}   & \multicolumn{2}{c}{$\name$}    \\ \midrule
\multirow{2}{*}{Bert-base} & LAN                    & \multicolumn{1}{c|}{53.17}    & \multicolumn{1}{c|}{124.81}    & \multicolumn{1}{c|}{43.98}  & \textbf{30.58}  & 1.44$\times$ \\ \cmidrule(l){2-7} 
                           & WAN                    & \multicolumn{1}{c|}{551.45}   & \multicolumn{1}{c|}{888.55}    & \multicolumn{1}{c|}{444.43} & \textbf{303.50} & 1.46$\times$ \\ \midrule
\multirow{2}{*}{GPT2-base} & LAN                    & \multicolumn{1}{c|}{39.60}    & \multicolumn{1}{c|}{57.57}     & \multicolumn{1}{c|}{38.33}  & \textbf{20.30}  & 1.89$\times$ \\ \cmidrule(l){2-7} 
                           & WAN                    & \multicolumn{1}{c|}{382.98}   & \multicolumn{1}{c|}{588.14}    & \multicolumn{1}{c|}{357.65} & \textbf{233.32} & 1.53$\times$ \\ \bottomrule
\end{tabular}
}
\end{table}

\begin{figure*}
    \centering
    \includegraphics[width=0.85\linewidth]{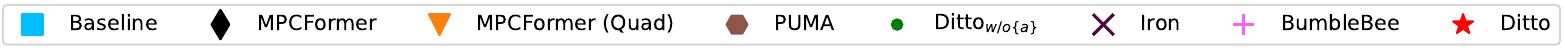}
    \subcaptionbox{Bert-base
    \label{fig:bert-base-2pc-efficiency}}[0.4\linewidth]
    {
        \includegraphics[width=1\linewidth]{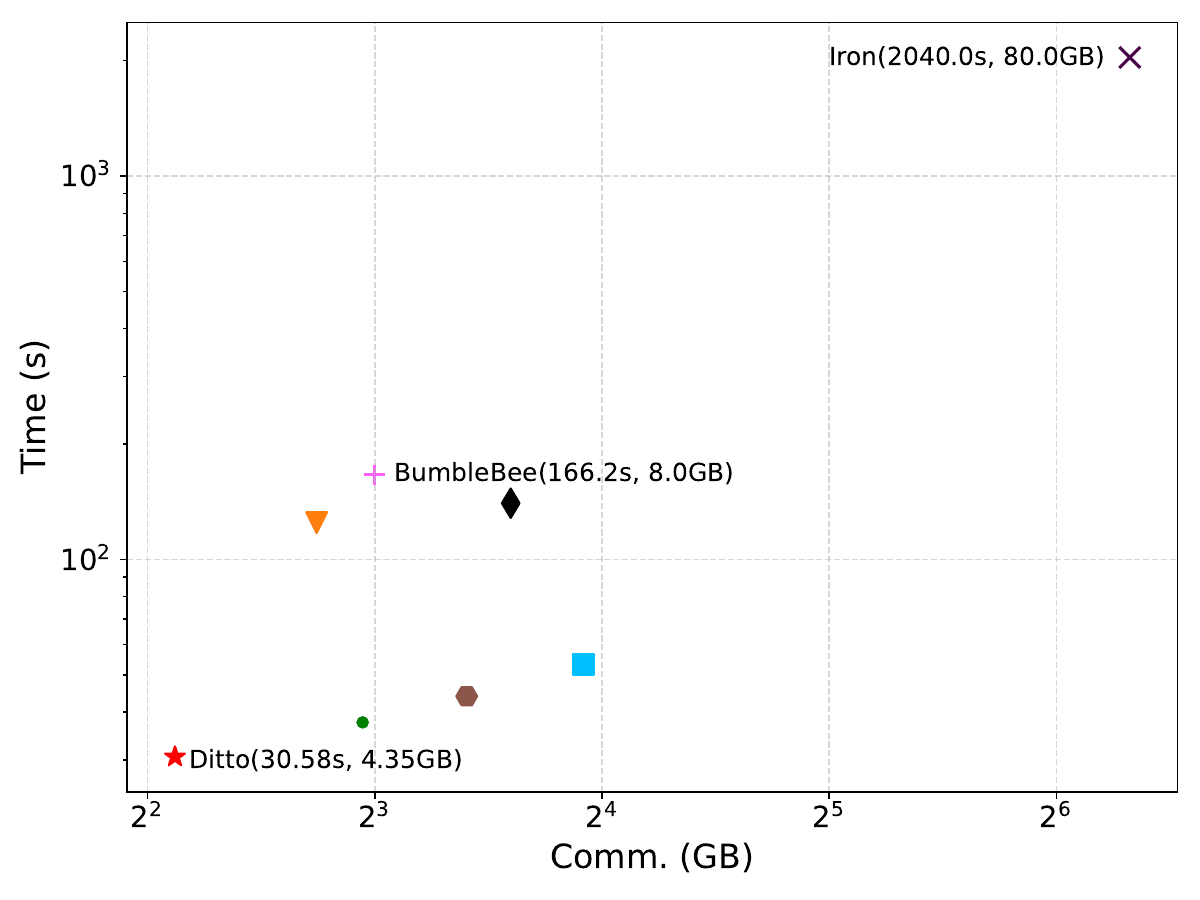}
    }
    \subcaptionbox{Bert-large
    \label{fig:bert-large-2pc-efficiency}}[0.4\linewidth]
    {
        \includegraphics[width=1\linewidth]{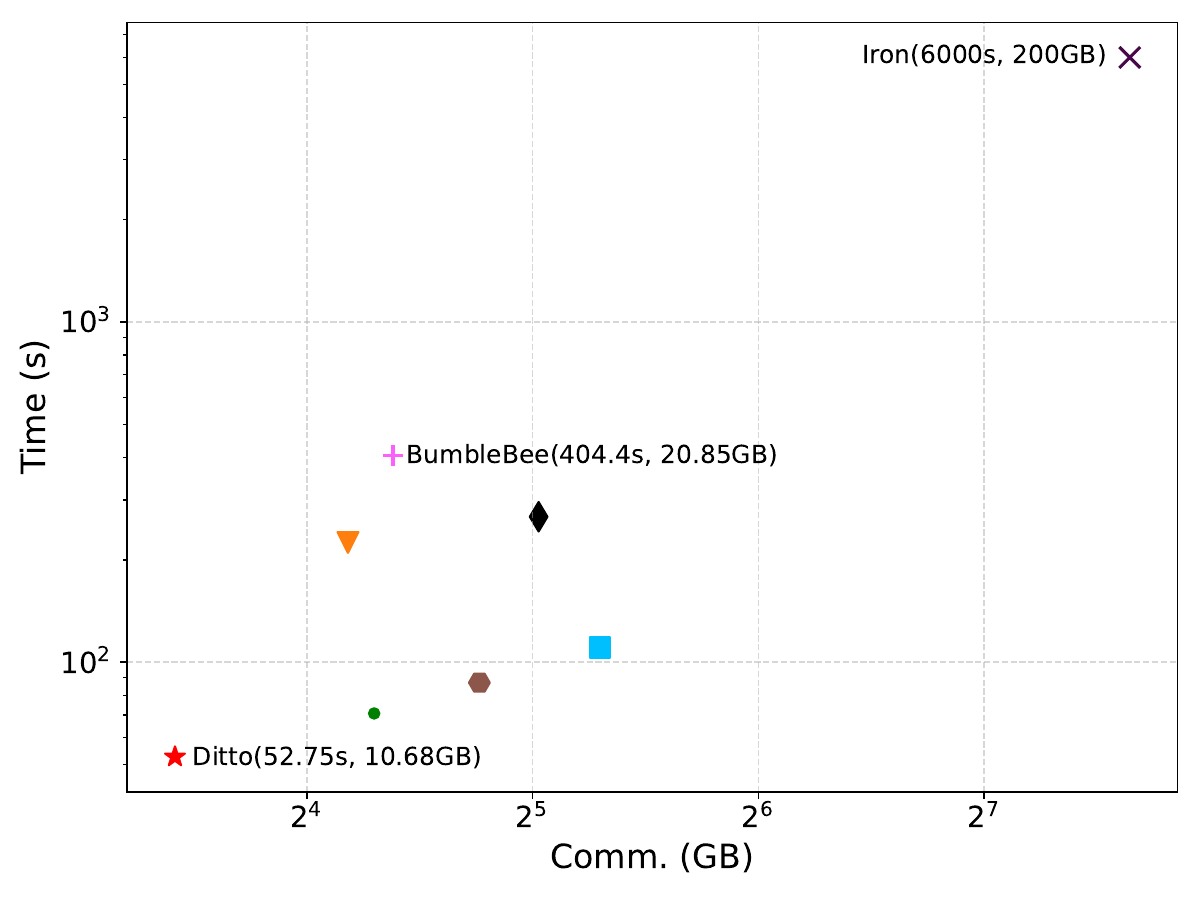}
    }
    \caption{Efficiency comparison to 2PC methods on Bert models. The input sequences are of length 128.
    }
    \label{fig:2pc-efficiency}
\end{figure*}

\paragraph{Comparison to 2PC works.}
We also evaluate the state-of-the-art 2PC works for secure Transformer inference to make a comprehensive comparison.
The concrete runtime and communication size are from the paper~~\cite{bumblebee}.
The efficiency results are illustrated in Figure~\ref{fig:2pc-efficiency}.
In general, \name is \textbf{orders of magnitude faster} than Iron~\cite{hao2022iron} and BumbleBee~\cite{bumblebee} on Bert (base and large) models with input sequence length of 128.

\section{Formulated Protocol Constructions}\label{append:protocols}
In this section, we present the formulated protocol constructions for the fixed-point computation of GeLU and Softmax functions mentioned in Section~\ref{sec:qa-functions}, and the type cast MPC primitives introduced in Section~\ref{sec:mpc-protocol}.

The approximation of GeLU function that computes $\mathsf{GeLU}(x) = 0.125x^2+0.25x+0.5$ is presented in Algorithm~\ref{protocol:gelu}.
The Softmax function that computes with different fixed-point representations are shown in Algorithm~\ref{protocol:softmax}.
\begin{algorithm*}[h]
\caption{Approximated $\mathsf{GeLU}$ Protocol}\label{protocol:gelu}
\begin{algorithmic}[1]
\REQUIRE
Fixed-point $x$ over $\texttt{FXP}_\ell^f$; Polynomial coefficients $a, b, c = \{0.125, 0.25, 0.5\}$
\ENSURE
Fixed-point $y$ over $\texttt{FXP}_\ell^f$;

\STATE $a_{int} = \lfloor a\cdot 2^f \rceil$, $b_{int} = \lfloor b\cdot 2^f \rceil$, $c_{int} = \lfloor c\cdot 2^f \rceil$
\STATE $\hat{y} = a_{int} \cdot x / 2^f + b_{int}, ~~~~~~\vartriangleright 0.125\cdot x + 0.25$
\STATE $y = x \cdot \hat{y} / 2^f + c_{int}, ~~~~~~\vartriangleright x\cdot(0.125\cdot x + 0.25) + 0.5$

\RETURN $y$

\end{algorithmic}
\end{algorithm*}

\begin{algorithm*}[h]
\caption{Approximated $\mathsf{Softmax}$ Protocol}\label{protocol:softmax}
\begin{algorithmic}[1]
\REQUIRE
Fixed-point $x$ over $\texttt{FXP}_\ell^f$; $\ell < \ell'$ and $f < f'$
\ENSURE
Fixed-point $y$ over $\texttt{FXP}_\ell^f$;

\STATE $x = x - \mathsf{Max}(x)$, ~~~~~~$\vartriangleright$ Max computes with precision bit $f$
\STATE $\hat{x} = \mathsf{Cast}(x, \texttt{FXP}_\ell^f, \texttt{FXP}_{\ell'}^{f'})$, ~~~~~~$\vartriangleright$ from $\texttt{FXP}_\ell^f$ to $\texttt{FXP}_{\ell'}^{f'}$
\STATE $\hat{x}_{exp} = \mathsf{Exp}(\hat{x})$, ~~~~~~$\vartriangleright$ Exponential computes with precision bit $f'$
\STATE $\hat{y} = \hat{x}_{exp} / \mathsf{Sum}(\hat{x}_{exp}, axis=-1)$
\STATE $y = \mathsf{Cast}(\hat{y}, \texttt{FXP}_{\ell'}^{f'}, \texttt{FXP}_\ell^f)$, ~~~~~~$\vartriangleright$ from $\texttt{FXP}_{\ell'}^{f'}$ to $\texttt{FXP}_\ell^f$
\RETURN $y$

\end{algorithmic}
\end{algorithm*}

In the following, the Algorithm~\ref{protocol:cast-down} depicts the construction of downcast protocol in the MPC domain.

\begin{algorithm*}[h]
\caption{Secure $\mathsf{DownCast}$ Protocol}\label{protocol:cast-down}
\begin{algorithmic}[1]
\REQUIRE
RSS-shared $\share{x}_{\ell}$ over $\texttt{FXP}_{\ell}^{f}$;
\ENSURE
RSS-shared $\share{x'}_{\ell'}$ over $\texttt{FXP}_{\ell'}^{f'}$, where $x/2^f=x'/2^{f'}$.

\STATE $P_i$ for $i \in \{0, 1, 2\}$ proceed as follows:
\begin{equation*}
\begin{split}
&x'_i = x_i \gg (f - f') \mod 2^{\ell'}\\
&x'_{i+1} = x_{i+1} \gg (f - f') \mod 2^{\ell'}, ~~~~~~\vartriangleright x/2^{f-f'} \mod 2^{\ell'} = x' \mod 2^{\ell'}\\
\end{split}
\end{equation*}

\RETURN$\share{x'}_{\ell'} = \{x'_0, x'_1, x'_2\}$.

\end{algorithmic}
\end{algorithm*}

\section{Correctness Analysis}\label{append:correctness}
In this section, we analyze the correctness of proposed type cast protocols in Section~\ref{sec:mpc-protocol}.
The type cast in MPC involves converting shares among rings of different sizes. Consider two rings, $\mathbb{Z}_{2^{\ell}}$ and $\mathbb{Z}_{2^{\ell'}}$, and let $\share{x}_\ell = \{x_0, x_1, x_2\}$ be the input sharing of $x$ over the first ring. Our goal is to obtain the sharing of $x$ over the second ring, denoted as $\share{x'}_{\ell'} = \{x'_0, x'_1, x'_2\}$. 
We note that $\share{x}_\ell$ and $\share{x}_{\ell'}$ are encoded over $\texttt{FXP}_\ell^f$ and $\texttt{FXP}_{\ell'}^{f'}$, respectively. Hence, we require $x/2^{f} = x'/2^{f'}$.
Note that to ensure correctness, we have the assumption that $x'$ can be represented using $\ell'$ bits, i.e., $x' \in [-2^{\ell'-1}, 2^{\ell'-1}-1]$.

\begin{proof}
Based on the relationship between $\{\ell, f\}$ and $\{\ell', f'\}$, we have two cases that correspond to downcast and upcast, respectively.

\emph{Case 1: $ \{\ell, f\} > \{\ell', f'\}$ (Downcast).} In Algorithm~\ref{protocol:cast-down}, the input $x$ is firstly right-shifted by $t = f - f'$ bits to lower the precision. The following step is to convert $x/2^{f-f'}$ to the smaller ring $\mathbb{Z}_{2^{\ell'}}$ using modulo operation.

The above steps can be formulated as 
\begin{equation}
\begin{split}
    x' = x/2^t &= ((x_0+x_1+x_2) \mod 2^\ell)/2^t \\
    &= x_0/2^t + x_1/ 2^t + x_2/ 2^t - w\cdot 2^{\ell-t} + w' \\
    &= (x_0/2^t + x_1/ 2^t + x_2/ 2^t - w\cdot 2^{\ell-t} + w') \mod 2^{\ell'} \\
    &= (x_0/2^t \mod 2^{\ell'}) + (x_1/ 2^t  \mod 2^{\ell'}) + (x_2/ 2^t  \mod 2^{\ell'}) \\
    &- (w\cdot 2^{\ell-t} \mod 2^{\ell'}) + w' \mod 2^{\ell'} \\
\end{split}
\end{equation}
where $w' \in \{0, 1, 2\}$ denotes the potential carry bits from the lower $t$ bits.  
Since we have $\ell-t = 64 - (18-8) = 54$, $\ell' = 32$, $w\cdot 2^{\ell-t} \mod 2^{\ell'} = 0$.
We can finally get 
\begin{equation}
\begin{split}
    x' &= (x_0/2^t \mod 2^{\ell'}) + (x_1/ 2^t  \mod 2^{\ell'}) + (x_2/ 2^t  \mod 2^{\ell'}) + w' \\
    &= (x'_0 + x'_1 + x'_2 + w') \mod 2^{\ell'}
\end{split}
\end{equation}
The probabilistic $w'$ occurs at the lowest significant bit, thus merely having a negligible impact of precision $2^{-f'}$.
The correctness of Algorithm~\ref{protocol:cast-down} thus holds.
\qed

\emph{Case 2: $ \{\ell, f\} < \{\ell', f'\}$ (Upcast).}
The input $x$ is firstly converted to the larger ring $\mathbb{Z}_{2^{\ell'}}$ using Algorithm~\ref{protocol:cast-up}, followed by a left-shifting of $t$ bits. 
Regarding Algorithm~\ref{protocol:cast-up}, the masking goes as 
\begin{equation}
    \begin{split}
        (x+r) \mod 2^\ell = x + r - \hat{w}\cdot2^\ell
    \end{split}
\end{equation}
where $\hat{w} = (x+r) \overset{?}{>} 2^\ell$. The above equation can be transformed into
\begin{equation}
    \begin{split}
        x \mod 2^{\ell'} = (x+r) \mod 2^\ell - r + \hat{w}\cdot2^\ell \mod 2^{\ell'}
    \end{split}
\end{equation}
The correctness holds as long as $\hat{w}$ is correct.
Recall that we add a bias to ensure that the MSB of $x$ is 0, $x+r$ wraps around $2^\ell$ if and only if the MSB of $r$ (i.e., $r_{\ell-1}$) is 1 and the MSB of $y = x +r \mod 2^\ell$ (i.e., $y_{\ell-1}$) is 0. Therefore, we can correctly compute the wrap as $\hat{w} = r_{\ell-1} \wedge \neg y_{\ell-1}$. As to the bias, we have a trick that limits the range of $x \in [-2^{\ell-2}, 2^{\ell-2}-1]$ and choose $2^{\ell-2}$ as the bias. As a result, any input $x = x + r \in [0, 2^{\ell-1}-1]$ is positive. After the conversion, the bias can be conveniently subtracted to eliminate its influence. The following left-shift operation can be regarded as multiplication by a public constant $2^t$, thus satisfying $x'/2^{f'} = x \cdot 2^t / 2^{f'} = x/2^f$.


\end{proof}


\section{Security Proof}\label{append:security-proof}

\begin{theorem}
Based on replicated secret sharing, the protocol ${\mathsf{DownCast}}$ securely performs the share extension against the semi-honest adversary, with honest-majority assumption.
\end{theorem}

\begin{proof}
The ${\mathsf{DownCast}}$ protocol relies on local right-shift and modulo operations, which are performed individually by each party on the shares they hold. No communication of shares between parties is required for these computations. Owing to the nature of the underlying RSS scheme, each party alone cannot reveal the secret data, proving the overall security of the protocol.
\end{proof}

\begin{theorem}
Based on replicated secret sharing, the protocol ${\mathsf{UpCast}}$ securely performs the share extension against the semi-honest adversary in the $(\mathsf{PRF}, \mathsf{DownCast})$-hybrid model, with honest-majority assumption.
\end{theorem}

\begin{proof}
${\mathsf{UpCast}}$ facilitates the computation by letting $P_2$ locally sample \textit{data-independent} correlated randomness and offload the subsequent computations to the left two parties, i.e., $P_0$ and $P_1$. Recall that we have honest-majority assumption, $P_2$ cannot collude with either $P_0$ or $P_1$. Hence, although $P_2$ knows the plaintext value of randomness, he cannot reveal the input $x$ without the information of the revealed $y$ in Step-5.
The randomness $r$ is a random $\ell$-bit integer. Its sharing over $\mathbb{Z}_{2^{\ell'}}$ is generated by $P_2$ and we let $P_0$ and $P_1$ use $\mathsf{DownCast}$ to obtain its sharing over $\mathbb{Z}_{2^{\ell}}$ in Step-3. As long as the security of $\mathsf{DownCast}$ holds, the security of Step-3 holds.
Subsequently, the mask-and-open operation computes $y=x+r$ over $\mathbb{Z}_{2^{\ell}}$. Since $r$ is uniformly sampled over $\mathbb{Z}_{2^{\ell}}$, and the computation modulos $2^\ell$, information-theoretical security is guaranteed. Despite the information of $y$, both $P_0$ and $P_1$ cannot crack $x$.
Regarding the computation of $\hat{w}$, it also merely involves local computations, thus leaking no information to help crack $x$.
Finally, in Step-6, the two random values $z_0, z_2$ over $\mathbb{Z}_{2^{\ell'}}$ generated using $\mathsf{PRF}$ in Step-2 are used to convert the two-out-of-two sharing of $x'$ into RSS.  $P_0$ and $P_1$ use $z_0$ and $z_2$ respectively to mask the shares they hold, while $P_2$ directly take $z_0$ and $z_2$ as his share. Since $z_0$ and $z_2$ are both uniformly sampled over $\mathbb{Z}_{2^{\ell'}}$, the masked shares exchanged between $P_0$ and $P_1$ does not leak any information. Since we assume the pair-wise seeds in $\mathsf{PRF}$ are securely distributed to the parties, the security of $\mathsf{PRF}$ holds, consequently the security of $\mathsf{UpCast}$.
\end{proof}

\section{Illustration of Training Loss During Distillation}\label{append:loss-distill}
\begin{figure}
    \centering
    \subcaptionbox{Training loss of Bert on CoLA dataset.
    \label{fig:dist-loss-bert}}[0.48\linewidth]
    {
        \includegraphics[width=\linewidth]{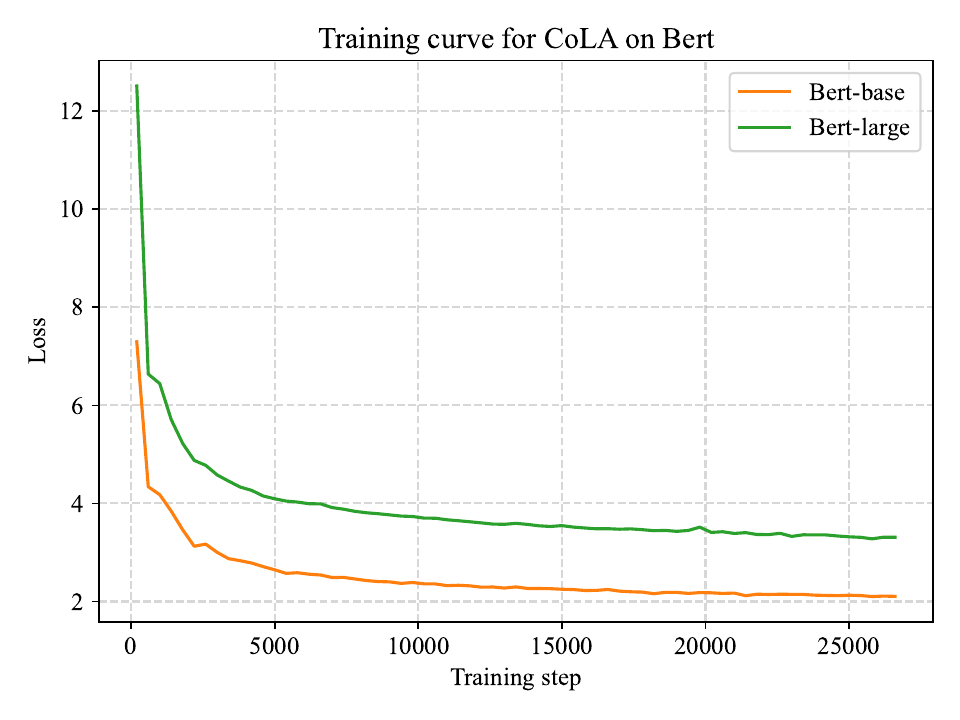}
    }
    \subcaptionbox{Training loss of GPT2 on Wikitext103 dataset.
    \label{fig:dist-loss-gpt2}}[0.48\linewidth]
    {
        \includegraphics[width=\linewidth]{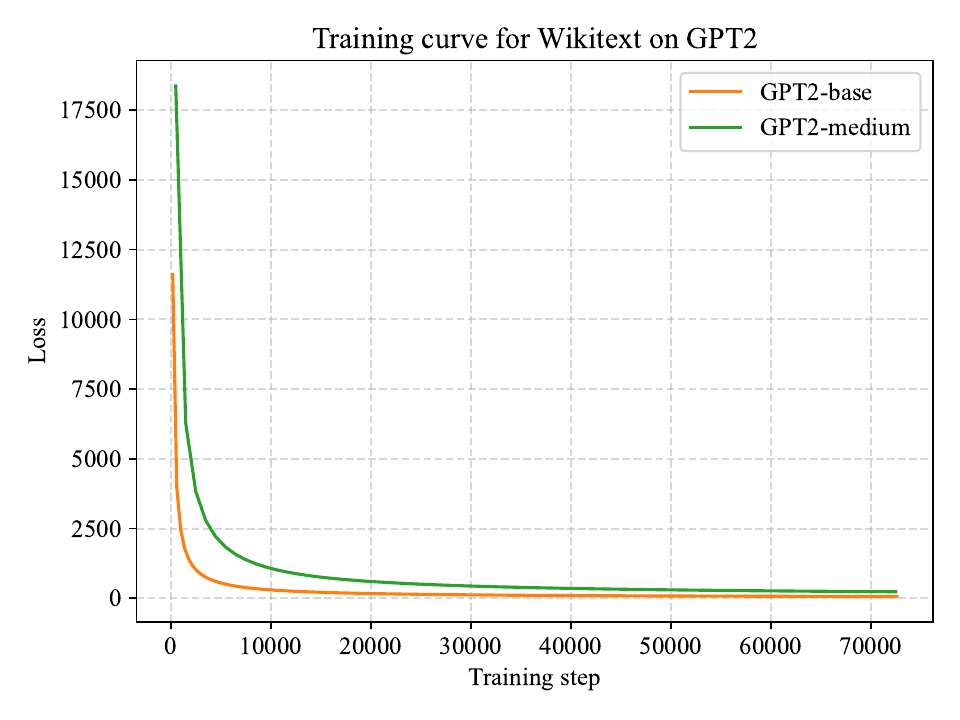}
    }
    \caption{Training loss of layer-wise outputs of Bert and GPT2 models.}
    \label{fig:distill-loss}
\end{figure}

The training curve in Figure~\ref{fig:distill-loss} depicts the loss between layer-wise outputs of the original model and the model generated by \name (with quantization and GeLU approximation). It is evident that the model produced by \name significantly deviates from the original model, with a maximum loss of 12 for Bert and nearly 18000 for GPT2. This substantial divergence indicates that without quantization-aware distillation, the converted model would have a low utility.

\section{Illustration of Activation Distribution}\label{append:distribution}
\begin{figure}[ht]
    \centering
    \subcaptionbox{Activation distribution of Bert-base
    \label{fig:dist-hs-bert-base}}[0.48\linewidth]
    {
        \includegraphics[width=\linewidth]{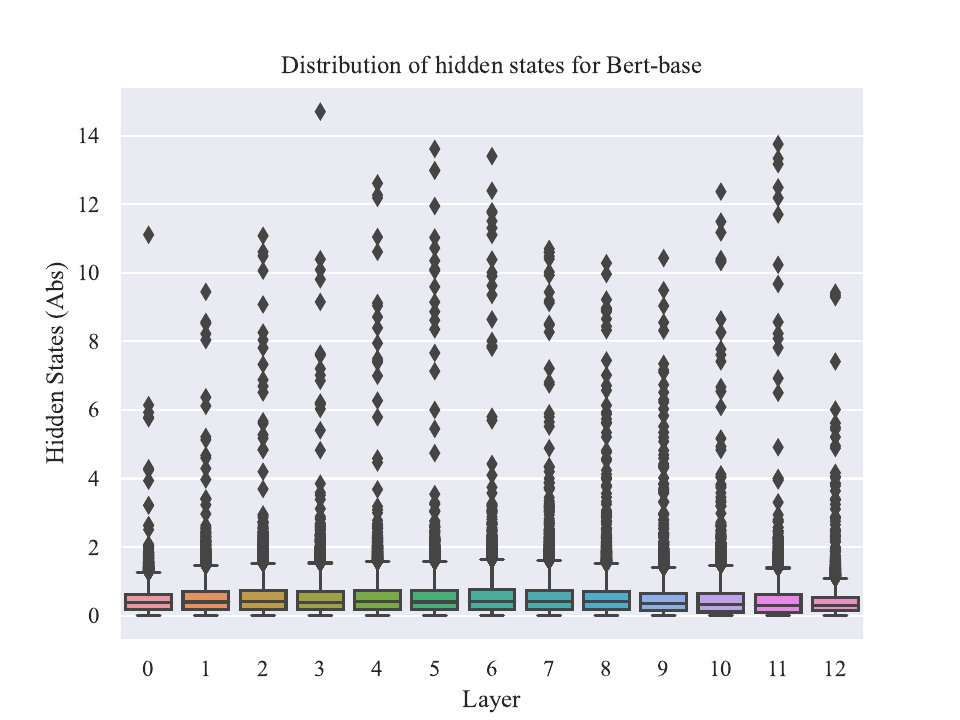}
    }
    \subcaptionbox{Activation distribution of Bert-large
    \label{fig:dist-hs-bert-large}}[0.48\linewidth]
    {
        \includegraphics[width=\linewidth]{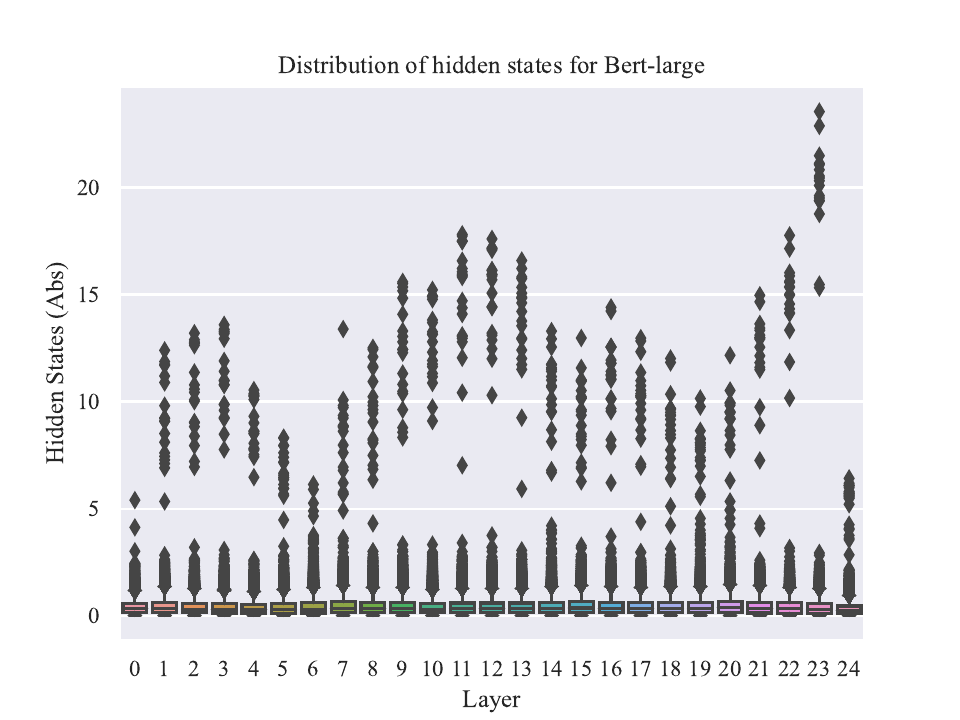}
    }
    \subcaptionbox{Activation distribution of GPT2-base
    \label{fig:dist-hs-gpt2-base}}[0.48\linewidth]
    {
        \includegraphics[width=\linewidth]{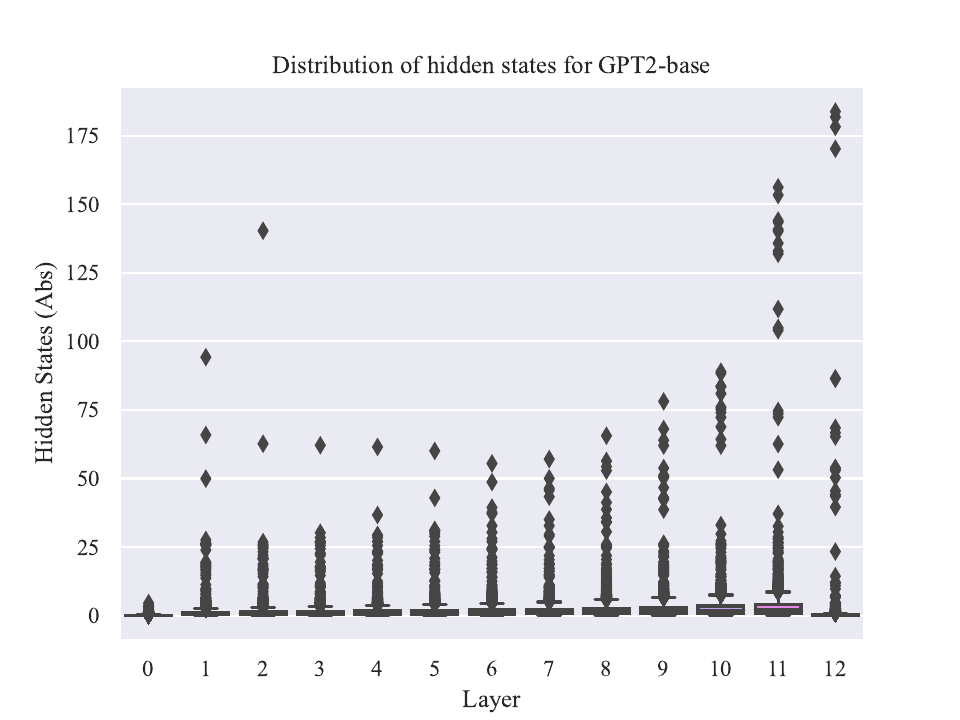}
    }
    \subcaptionbox{Activation distribution of GPT2-medium
    \label{fig:dist-hs-gpt2-medium}}[0.48\linewidth]
    {
        \includegraphics[width=\linewidth]{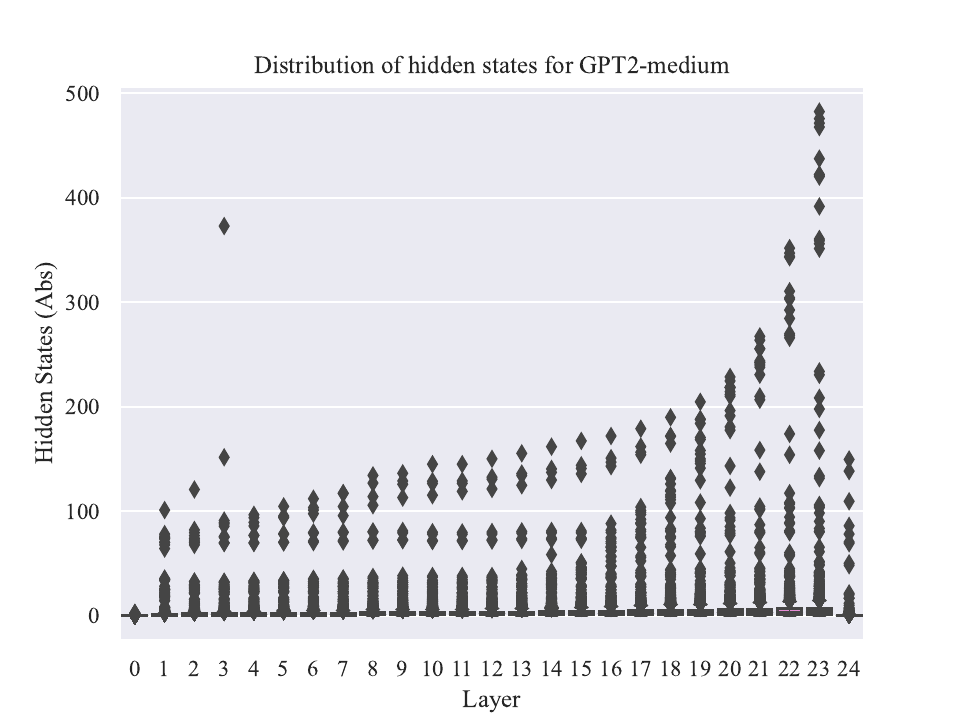}
    }
    \caption{Activation distribution on Bert and GPT2 models.}
    \label{fig:dist}
\end{figure}

In this section, we analyze the activation distribution of Bert and GPT2 models, focusing on the hidden states generated by the intermediate Transformer blocks. As depicted in Figure~\ref{fig:dist}, we observe that the majority of activations in these intermediate layers have absolute values close to zero, with only a small proportion of outliers. For Bert, the outliers fall below 25, while for GPT2, they are below 500. This distribution signifies that the quantization scheme employed in \name is capable of representing all intermediate values without encountering significant overflows.

\section{Illustration of Fixed-point Inference}\label{append:fixed-point-inference}
\begin{figure}
    \centering
    \includegraphics[scale=0.45]{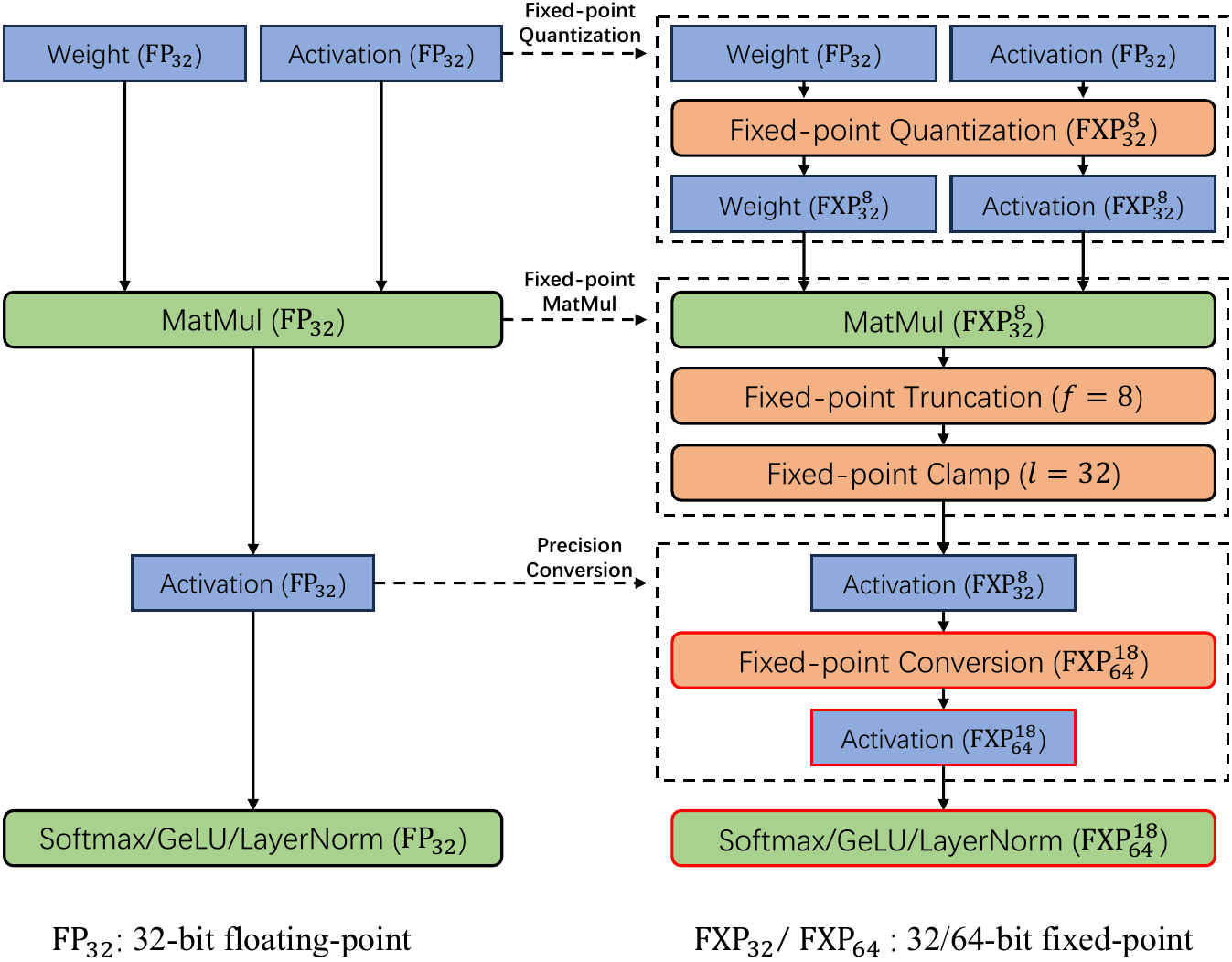}
    \caption{Comparison of fixed-point quantization schemes against original floating-point scheme.}
    \label{fig:comparison}
\end{figure}

The difference between fixed-point inference against traditional floating-point inference is illustrated in Figure~\ref{fig:comparison}. FP32 denotes float32, and FXP$_\ell^f$ represents $\ell$-bit fixed-point number with $f$-bit precision. We have $\ell \in \{32, 64\}, f \in \{8, 18\}$.
In floating-point inference (the \textbf{left} part), all the computations are computed using FP32. 
While in \name (the \textbf{right} part), all the variables (i.e., activations and weights) in each layer are quantized into fixed-point representation with different precision (marked in {Orange}). 
The concrete computations are also carried out using fixed-point arithmetic.

For the linear layers (using FXP$_{32}^{8}$), the fixed-point weights and activations serve as the inputs to linear operation, and the outputs are \textbf{truncated and clipped} to align the fixed-point representation.

While for the non-linear layers, we first perform \textbf{fixed-point conversion} to raise the precision, i.e., from $2^{-8}$ to $2^{-18}$ for the sake of numerical stability. The non-linear functions are approximated using fixed-point arithmetic (using FXP$_{64}^{18}$) that are detailed in Section~\ref{sec:fixed-point-arithmetic}.

\section{Hyper-parameter Choice}\label{append:hyperparameter}
\paragraph{Fine-tuning Configuration.}
Regarding the fine-tuning of Bert models for different classification tasks (re-train the last prediction layer), we use a batch size of 32 for Bert-base and 16 for Bert-large. All the inputs are of sequence length 128.
We train the models for 3 epochs on RTE, CoLA, QQP and QNLI datasets. We run a grid search with learning rate in [2e-5, 3e-5, 4e-5, 5e-5]. While for GPT2 models, we reuse the trained model parameters from Hugging Face.
Concretely, we use pre-trained GPT2-base~\footnote{GPT2-base on wikitext-103: \url{https://huggingface.co/Graphcore/gpt2-wikitext-103}}
and GPT2-medium
~\footnote{GPT2-medium on wikitext-103\url{https://huggingface.co/Graphcore/gpt2-medium-wikitext-103}}
models pre-trained over the Wikitext-103 dataset~\cite{merity2016pointer}.

\paragraph{Distillation Configuration.}
The distillation involves two stages: the hidden state distillation and logits distillation, with learning rates of 5e-5 and 1e-5, respectively. In general, we follow the hyper-parameter setting in \cite{li2023mpcformer}. 
Regarding Bert models, we train the student model for 50 epochs on RTE, 50 epochs on CoLA, 5 epochs on QQP and 10 epochs on QNLI. All the input sequences are of length 128.
For GPT2 models, we train for 1 epoch on Wikitext-103, and the input sequences are of length 50.
For the Bert-base model, we use a batch size of 32. While for Bert-large and GPT2 models, we use a batch size of 16 due to GPU memory limitation.


\section{Polynomial approximation of GeLU in PUMA}\label{append:poly-gelu-in-puma}
PUMA~\cite{puma-2023} proposed to use a piece-wise approximation of low-degree polynomials for more efficient yet accurate computation of secure GeLU function. In general, the GeLU approximation is split into four splines as follows:
\begin{equation}\label{eq:gelu-approx}
\mathsf{GeLU}(x)=
\begin{cases}
0, & x<-4 \\
f_0(x), & -4 \le x < -1.95 \\
f_1(x), & -1.95 \le x \le 3 \\
x, & x >3
\end{cases},
\end{equation}
where the polynomials $f_0, f_1$ are obtained using $\mathsf{numpy.ployfit}$\footnote{\url{https://numpy.org/doc/stable/reference/generated/numpy.polyfit.html}}. The coefficients of the two polynomials are listed below.

\begin{equation}\label{eq:f0f1}
\begin{cases}
f_0(x) &= -0.011034134030615728 x^3 -0.11807612951181953 x^2 \\
&- 0.42226581151983866 x -0.5054031199708174\\
f_1(x) &= 0.0018067462606141187x^6 -0.037688200365904236 x^4 \\
&+ 0.3603292692789629x^2 + 0.5x + 0.008526321541038084
\end{cases}
\end{equation}

\section{Tailored Exponential Approximation}\label{append:approximation}
We here describe the tailored exponential approximation for softmax~\cite{puma-2023} that is utilized in this work.
In general, the exponential function is approximated using a two-segment piecewise function defined in Equation~\ref{eq:negexp}. The approximation is specifically designed for the softmax function, which normalizes the input by subtracting the maximum value ($x = x - \mathsf{max}_i(x)$). As a result, the inputs to the exponential function are non-positive, allowing us to employ a Taylor series approximation that maintains precision without significant loss.
In our implementation, we set the parameters $T_{\mathrm{exp}}$ to $-14$ and $t$ to $5$ to ensure an efficient and accurate approximation of the exponential function. The specific details and formulas can be found in \cite{puma-2023}.

\begin{equation}\label{eq:negexp}
\mathsf{Exp}(x) = \begin{cases}
    0, &x < T_{\exp} \\
    (1+\frac{x}{2^t})^{2^t}, &x\in [T_{\exp},0].
\end{cases}
\end{equation}

\end{document}